\documentclass[useAMS,usenatbib,twocolumn]{mn2e}
\pdfoutput=1
\usepackage{amssymb}
\usepackage{amsmath}
\usepackage{natbib}
\usepackage[pdftex]{graphicx}
\usepackage{subfigure}
\usepackage{booktabs}
\begin{document}

\title[Tidal instability]
{Nonlinear evolution of the tidal elliptical instability in gaseous planets and stars}
      \author[A.J. Barker \& Y. Lithwick]{Adrian J. Barker\thanks{E-mail:
	  adrianjohnbarker@gmail.com} and Yoram Lithwick \\
	  Center for Interdisciplinary Exploration and Research in Astrophysics (CIERA) \& \\ Dept. of Physics and Astronomy, Northwestern University, 2145 Sheridan Rd, Evanston, IL 60208, USA.}
	
\pagerange{\pageref{firstpage}--\pageref{lastpage}} \pubyear{2013}

\maketitle

\label{firstpage}

\begin{abstract}
Tidally distorted rotating stars and gaseous planets are subject to a well-known linear fluid instability---the elliptical instability.
It has been proposed that this instability might drive enough energy dissipation to
 solve the long-standing problem of the origin of tidal dissipation in stars and planets. 
But the nonlinear outcome of the elliptical instability has yet to be investigated in the parameter regime of interest, and the resulting turbulent energy dissipation has not yet been quantified. We do so by performing three dimensional hydrodynamical simulations of a small patch of a tidally deformed fluid planet or star subject to the elliptical
instability. We show that when the tidal deformation is weak, the nonlinear outcome of the instability leads to the formation 
of long-lived columnar vortices aligned with the axis of rotation. These vortices shut off the elliptical instability, and the net 
result is insufficient energy dissipation to account for tidal dissipation. However, further work is required to account for effects neglected here, including magnetic fields, turbulent convection, and realistic boundary conditions.
\end{abstract}


\section{Introduction}

Identifying and elucidating the mechanisms of tidal dissipation in fluid bodies, such as giant planets and stars, remains an important unsolved problem. Tidal dissipation is thought to be responsible for several different phenomena in close binary systems composed of one or more fluid bodies, including: synchronisation of the spins and circularisation of the orbits of close binary stars (e.g.~\citealt{Meibom2005}; \citealt{Mazeh2008}), synchronisation and circularisation, and tidal heating, of short-period extrasolar planets (e.g.~\citealt{BLM2001}; \citealt{Husnoo2012}), the formation and maintenance of the Laplace resonance amongst the Galilean satellites of Jupiter \citep{YoderPeale1981}, and possibly also playing a role in the spin-orbit alignment \citep{Albrecht2012} and orbital evolution of hot Jupiters \citep{Jackson2009}.

The efficiency of these evolutionary processes is often parametrized by the dimensionless tidal quality factor $Q$ associated with each body, which is an inverse measure of their dissipative properties, and is analogous to the corresponding quantity for a forced, damped, oscillator \citep{GoldSot1966,MurrayDermott1999}. We define the function 
$
Q^{\prime} = \frac{3}{2 k_{2}}\frac{2\pi E_{0}}{\int |\dot{E}| dt}
$ 
\citep{Gio2007},
which is the modified tidal quality factor, where $k_{2}$ is the Love number, being a measure of the degree of central condensation of the body ($k_{2}=3/2$ for a homogeneous body), $\dot{E}$ is the rate of dissipation of energy, and $E_{0}$ is the maximum potential energy in the tidal deformation. Understanding the efficiency of tidal dissipation in stars and planets requires identifying the mechanisms that contribute to $Q^{\prime}$, and their dependence on the tidal frequency, the amplitude of the tide, the interior structure of the body, its (differential) rotation rate, and possibly the past history of the system.

The inferred values of $Q^{\prime}$ required to explain observations depend on the particular problem in which tides are invoked as the explanation. Arguably, two of the most difficult constraints to be satisfied by theory are the observationally inferred circularisation periods of solar-type close binary stars, and the required magnitude of Jupiter's tidal dissipation. The former seems to require $Q^{\prime}\sim 10^{5}$ for close-binary stars, if their orbits are circularised over their main--sequence lifetimes due to tides (\citealt{Meibom2005}; \citealt{Gio2007}). The magnitude of Jupiter's dissipation, required to explain the formation of the Laplace resonance through differential tidal expansion of the orbits of the Galilean satellites \citep{YoderPeale1981}, also typically implies $Q^{\prime} \sim 10^{5}$. Astrometric observations of the Galilean moons have since indicated a similar magnitude \citep{Lainey2009}. The dissipation inside the planet required to explain the circularisation of hot Jupiters on short-period orbits has also been estimated to imply $Q^{\prime} \sim 10^{6}$ (\citealt{WuMurray2003}, \citealt{JacksonI2008}). The similarity in these observationally inferred values for the stellar and planetary dissipation might be misleading, since it is very likely that the internal structure of the body plays an important role in determining the efficiency of dissipation, as may also the amplitude of the tides, which are very different for each of these problems.

The tidal response in a primary fluid body is often decomposed into two components. One is a quasi-hydrostatic prolate spheroidal bulge that follows the motion of the companion. This is referred to as the equilibrium tide, and is the divergence-free Lagrangian displacement field $\boldsymbol{\xi}^{e}$ with (e.g.~\citealt{Goldreich1989})
\begin{eqnarray}
\label{eqmtideeqns}
\xi^{e}_{r} = -\frac{\Psi}{g},
\end{eqnarray}
where $\Psi$ is the tidal
gravitational potential (neglecting self-gravity) experienced by the body and $g$ is the
gravitational acceleration. The equilibrium tide is dissipated through its action with turbulent convection \citep{Zahn1966}, though the efficiency of that process is reduced when the tidal frequency exceeds the turnover frequency of convective eddies, and seems unlikely to explain the observations in this regime (e.g.~\citealt{Goldreich1977}; \citealt{Penev2007}; \citealt{OgilvieLesur2012}).

The remaining part of the response is dubbed the dynamical tide, and understanding this component involves the study of the excitation, propagation and dissipation of waves excited (indirectly) by the tidal potential. These waves are of low frequency compared with the dynamical frequency, and are therefore approximately incompressible, being restored by buoyancy and/or rotation.

The dynamical tide can driven by two different mechanisms. The first is forcing of the dynamical tide by the equilibrium tide through the linear inertial terms. This mechanism has received the most attention (e.g.~\citealt{Zahn1975}; \citealt{Savonije1995}; \citealt{Gio2004}; \citealt{Wu2005b}; \citealt{Gio2007}, \citealt{IvanovPap2007}, \citealt{GoodmanLackner2009}). However, these studies are linear, and they cannot capture any excitation of the dynamical tide through instabilities of the equilibrium tidal flow.  If, on the other hand, the equilibrium tidal flow becomes unstable to parametric instabilities, the resulting growth of small-scale internal waves and their saturation by secondary instabilities could result in turbulence, and secular energy dissipation. This mechanism is entirely distinct from the linear tidal forcing of the dynamical tide, and has so far received much less attention because it involves nonlinear processes, requiring numerical simulations to quantify. It is the purpose of this paper to begin to understand and quantify the tidal dissipation resulting from instabilities of the equilibrium tide\footnote{\cite{Weinberg2012} has considered the parametric instability of the equilibrium tide, leading to the excitation of g-modes, in a non-rotating solar-type star, finding that such a mechanism is probably not important for tidal dissipation. Here we study the instability of the equilibrium tide in a rotating fluid body, and its nonlinear evolution.}.

The equilibrium flow in a rotating tidally deformed body has elliptical streamlines, which have their semi-major axes aligned with the companion. 
In the absence of stabilising buoyancy forces, as would be relevant in the convective regions of a fluid planet or star,
it is well known that any ideal flow
 with elliptical streamlines is unstable to parametric instabilities, which excite inertial waves due to the periodic variation of the angular velocity along a streamline \citep{Pierrehumbert1986,Bayly1986,Kerswell2002}.  This is the so-called elliptical instability.

The elliptical instability has been proposed to be of significant importance for tidal dissipation in binary stars by \cite{Rieutord2004}. Since then, some remarkable experiments have been performed by \cite{Lacaze2004} as well as \cite{LeBars2007,LeBars2010} to study the instability in the laboratory, in which fluid is contained within a deformable elastic boundary, which is rotated at a certain speed. These experiments generally find that the resulting turbulence is potentially important for tidal dissipation in terrestrial planet cores.  In this paper, we study the dissipation in astrophysical fluid bodies using an idealised local model, and our results indicate that it is probably less important in explaining astrophysical observations than it has been previously suggested by \cite{Rieutord2004}.

We study the nonlinear evolution of the elliptical instability in a small patch of a rotating tidally deformed fluid body\footnote{Any local Cartesian model necessarily neglects the geometrical effects that alter the spatial structure of the inertial wave response in a spherical or spherical shell geometry (e.g.~\citealt{Gio2004,Wu2005b,IvanovPap2007}). The results of this simplification are difficult to determine without performing detailed global simulations, which we defer to future work.}. Our primary method involves high resolution numerical simulations in a local Cartesian model, performed using spectral methods. First we outline the relevant background to motivate and understand our results in the rest of \S 1. A simple model of the dissipation resulting from the elliptical instability is presented in \S 2, together with its potential astrophysical relevance, followed by a description of our model in \S 4--6. Our results are described in \S 7, followed by a discussion and conclusion in \S 8--9, in which the astrophysical implications of this work are discussed, together with various extensions which can build upon this work.

\subsection{Elliptical instability and the properties of inertial waves}
The elliptical instability is a generic linear instability that occurs when fluid flows on elliptical streamlines, which occurs when a rotating fluid body is subject to a tidal deformation. The instability is driven parametrically by the periodic variation of the fluid velocity along an elliptical streamline (see the review by \citealt{Kerswell2002}). An initially two-dimensional elliptical vortex can become unstable to this inherently three-dimensional instability, which takes the form of a parametric resonance with pairs of inertial waves that couple with the elliptical strain field. 

Inertial waves exist in rotating fluids, and are restored by the Coriolis force. In a homogeneous fluid rotating at the uniform rate $\Omega$ about $\boldsymbol{e}_{z}$, they satisfy the dispersion relation\footnote{A single plane inertial wave is an exact solution of the incompressible Euler equations in a rotating frame. However, these waves are generically unstable to parametric instabilities (e.g.~\citealt{LifshitzFabijonas1996}; \citealt{Miyazaki1998}). This is because they undergo resonant triad interactions with pairs of secondary waves, whose growth rates increase linearly with the ratio of wave velocity amplitude to the phase velocity. The waves are expected to rapidly break as a result of these secondary instabilities when this ratio is somewhat smaller than unity. These instabilities have been recently observed in the laboratory by \cite{Dauxois2011}.}
\begin{eqnarray}
\label{dispersion}
\omega = s\frac{2\boldsymbol{\Omega}\cdot\boldsymbol{k}}{k} =  s\frac{2\Omega k_{z}}{k} = 2\Omega s\cos \theta,
\end{eqnarray}
where $s = \pm 1$, $k = |\boldsymbol{k}| = \sqrt{k_{h}^{2}+k_{z}^{2}}$ is the wavenumber and $k_{z}$ ($k_{h}$) is the component along (perpendicular to) the rotation axis, with $\theta$ being the angle that the wavevector makes with the vertical. Note that this is independent of $k$, so that inertial waves can exist on all spatial scales, in the absence of viscosity. These waves exist whenever $|\omega| \leq 2|\Omega|$. The fluid particles in the wave move in anticyclonic circles about $\boldsymbol{k}$, in a plane inclined to the horizontal by an angle $\theta$. These motions are equivalent to pure epicyclic motions in horizontal planes if $\theta = 0$. 

To illustrate the mechanism of instability, consider the case when the ellipticity is very small.  Then, the streamlines are nearly circular, 
and the fluid supports the usual inertial waves, but the properties of the waves are modified
slightly by the ellipticity. In particular, the elliptical deformation of the vortex can be considered as a wave
of wavenumber $k\rightarrow 0$ and
 of frequency $2\gamma$, where $\gamma$ is the rotation rate \citep{Goodman1993}. This disturbance can nonlinearly interact with each of two initially infinitesimal amplitude propagating waves with ($\pm \omega, \mathbf{k}$), which produces disturbances with ($\pm \omega \pm 2\gamma, \mathbf{k}$). If $\omega = \mp\gamma$, two of these four possible components will also be inertial waves which satisfy Eq.~\ref{dispersion}. Since the two waves that are excited have $\omega = \pm \gamma$, this mechanism reinforces the original infinitesimal amplitude waves at first order in the wave amplitude. Therefore, the amplitudes of the two waves will grow exponentially, if this coherent driving of the waves is maintained.

\begin{figure}
  \begin{center}
     \subfigure{\includegraphics[trim=0cm 0cm 0cm 0cm, clip=true,width=0.48\textwidth]{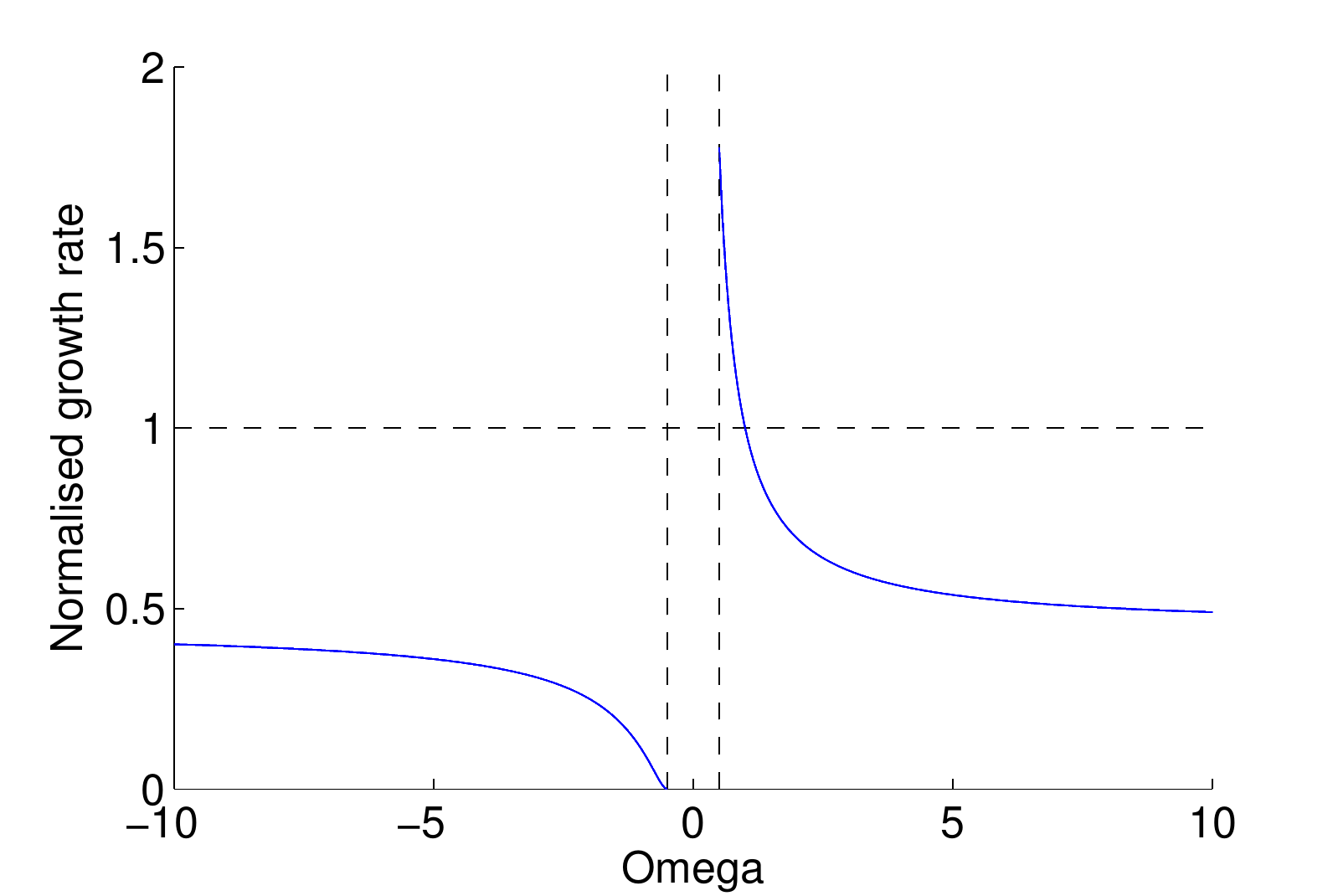} } 
    \end{center}
  \caption{Maximum growth rate of the elliptical instability as a function of the rotation of the fluid $\Omega$, normalised to the growth rate when $n=0$. The growth rate is generally modified by approximately a factor of 2, except for the appearance of a range in which the instability is inoperable for $\Omega\in (-0.5,0.5)$ (i.e. $n\in(-1.5,-0.5)$).}
  \label{rotgrowth}
\end{figure}

In this paper we are interested in a local model of a tidally deformed fluid body. In this local approximation, the elliptical instability takes its simplest form, and is absent of boundary effects. The growth rate of these linear instabilities scales with the ellipticity, and for the case of an unbounded elliptical vortex, the maximum growth rate is \citep{Waleffe1990}
\begin{eqnarray}
\label{maxgr}
\sigma = \frac{9}{16}\epsilon \gamma,
\end{eqnarray}
for waves with the appropriate phase and angle to align with the direction of maximum strain, where $\epsilon$ is the ellipticity and $\gamma$ is the rotation rate of the vortex. When the elliptical bulge is also rotating at the rate $n$, and the fluid is rotating at the rate $\Omega$, and $\gamma = \Omega -n$, the maximum growth rate is \citep{Craik1989}
\begin{eqnarray}
\label{maxgrot}
\sigma = \frac{9}{16}\epsilon \gamma \frac{\left(3\gamma+2n\right)^{2}}{9\left(\gamma+n \right)^{2}}.
\end{eqnarray}
We plot this quantity as a function of $\Omega$, normalised to the growth rate in Eq.~\ref{maxgr}, by taking $\gamma =1$, so that $n$ is changed by modifying $\Omega$, in Fig.~\ref{rotgrowth}. The addition of this rotation has the sole effect of modifying the frequencies of the most unstable waves, and this results in the creation of a range in $\Omega$ at which the elliptical instability is inoperable, when $\Omega \in (-0.5,0.5)$, located inside the vertical dashed lines on the figure. Outside of this range, the growth rate is usually decreased by an $O(1)$ factor by the rotation of the deformation, except in the range $\Omega \in[0.5,1)$, where the growth rate is slightly larger than the nonrotating case.

Understanding the nonlinear interactions between inertial waves is essential for understanding the dynamics of rotating turbulence with small Rossby number Ro (the ratio of inertial to Coriolis accelerations). For reasons that will become clear later in this paper, we briefly review some of  the properties of rotating turbulence. The only type of steady flow that can exist in the presence of rapid rotation is one with angular velocity that is constant on cylinders aligned with the axis of rotation, which is the Proudman-Taylor theorem, often written $\boldsymbol{\Omega}\cdot \nabla\boldsymbol{u} = 0$ (\citealt{Hough1897}; \citealt{Proudman1916}; \citealt{Taylor1917}). Homogeneous rotating turbulence starting with isotropic initial conditions with Ro $ \ll 1$ is found to approximately two-dimensionalise, developing coherent vortical structures elongated along the axis of rotation. This is thought primarily to result from nonlinear interactions between inertial waves, which can produce columnar structures indirectly (e.g.~\citealt{Waleffe1993}; \citealt{SmithWaleffe1999}). This is because nonlinear interactions between inertial waves tend to transfer energy to waves with smaller $k_{z}/k$ i.e. to larger vertical scales and smaller frequencies. In addition, the forward energy cascade is found to be inhibited by the Coriolis force, so that the dissipation rate is reduced (e.g.~\citealt{Bardina1985}; \citealt{Jacquin1990}). 

\section{Elliptical instability: order of magnitude estimate}
\label{orderofmag}

To illustrate the potential importance of the elliptical instability, and to justify the relevance of this work, we provide a simple order of magnitude estimate for the resulting tidal dissipation. We consider the tidal flow in body 1 (the primary) of mass $m_{1}$ and radius $R_{1}$, induced by the tidal potential due to body 2 (the secondary) of mass $m_{2}$. For the synchronisation problem, the rotation of the primary $\Omega$ is not equal to the orbital mean motion of the secondary $n$, which we assume is on a circular orbit in the equatorial plane of the primary. 
The elliptical instability of the equilibrium tidal flow has a growth rate (when $\epsilon \ll 1$) scaling as
\begin{eqnarray}
\frac{1}{t_{grow}} \sim \epsilon \gamma,
\end{eqnarray}
where $\gamma=|\Omega-n|$ (and $\epsilon \sim |\xi^{e}_{r}(R_{1})|/R_{1}$). We assume the unstable mode to have spatial scale $l$, with velocity amplitude $u$, and to decay due to secondary instabilities at the rate
\begin{eqnarray}
\frac{1}{t_{damp}} \sim \frac{u}{l}.
\end{eqnarray}
In this picture, a steady state is reached when $t_{grow} \sim t_{damp}$, at which the velocity of the mode should saturate.
The corresponding dissipation rate, assuming such a continual energy transfer to the mode, is
\begin{eqnarray}
D \sim m_{1}u^{2}/t_{damp} \sim m_{1}l^{2} \epsilon^{3}\gamma^{3} .
\end{eqnarray}
The energy stored in the equilibrium tide response is 
\begin{eqnarray}
E_{0} \sim \frac{Gm_{1}^{2}}{R_{1}}\epsilon^{2}.
\end{eqnarray}
In this case,
\begin{eqnarray}
\epsilon \sim \frac{m_{2}}{m_{1}}\left(\frac{R_{1}}{a}\right)^{3} \sim \left(\frac{m_{2}}{m_{1}+m_{2}}\right) \left(\frac{P_{dyn}}{P}\right)^{2},
\end{eqnarray}
where we have defined $P_{dyn}=2\pi \sqrt{\frac{R_{1}^{3}}{Gm_{1}}}$.
Therefore the tidal quality factor that results, taking $l \sim R_{1}$, and $\gamma=2\pi/P$, is
\begin{eqnarray}
\label{ordermagq}
Q^{\prime} &\sim & \frac{\gamma E_{0}}{D} \sim\left(\frac{m_{1}+m_{2}}{m_{2}}\right) \left(\frac{P}{P_{dyn}}\right)^{4} \\
&\approx & 4.4 \times 10^{5} \left(\frac{m_{1}+m_{2}}{m_{2}}\right)\left(\frac{2.8 \; \mathrm{hrs}}{P_{dyn}}\right)^{4}\left(\frac{P}{3 \; \mathrm{d}}\right)^{4}.
\end{eqnarray}

To illustrate the importance of this dissipation, the timescale over which an initially eccentric hot Jupiter is circularised (e.g.~\citealt{GoldSot1966}) is
\begin{eqnarray}
\nonumber
\tau_{e} &=&\frac{4}{63}\frac{Q^{\prime}}{2\pi} \frac{m_{1}}{m_{2}}\left(\frac{m_{1}+m_{2}}{m_{1}}\right)^{\frac{5}{3}} \frac{P^{\frac{13}{3}}}{P_{dyn}^{\frac{10}{3}}}
\\ &\approx& 5 \; \textrm{Gyr} \left(\frac{2.8\; \mathrm{hr}}{P_{dyn}}\right)^{\frac{22}{3}} \left(\frac{P}{4.5\; \mathrm{d}}\right)^{\frac{25}{3}},
\label{tecc}
\end{eqnarray}
where we have taken $m_{1}=10^{-3}m_{2}$ in the last expression. Similarly, the timescale to synchronise its spin is
\begin{eqnarray}
\nonumber
\tau_{\Omega} &=& \frac{4}{9}\frac{Q^{\prime} r_{g}^{2}}{2\pi} \left(\frac{m_{1}+m_{2}}{m_{2}}\right)^{2} \frac{P^{4}}{P_{rot} P_{dyn}^{2}} \\ \nonumber &\approx&
 5 \; \textrm{Gyr} \left(\frac{2.8\; \mathrm{hr}}{P_{dyn}}\right)^{6} \left(\frac{0.5\; \mathrm{d}}{P_{rot}}\right) \left(\frac{P}{11.5 \; \mathrm{d}}\right)^{8},
 \label{tsync}
\end{eqnarray}
where $r^{2}_{g}=0.254$ is the squared dimensionless radius of gyration for Jupiter, and we have taken $m_{1}=10^{-3}m_{2}$ in the last expression. Note the strong orbital period dependence of these evolutionary timescales, which is stronger than the case with a constant $Q^{\prime}$, arises because this is a nonlinear mechanism of tidal dissipation.

These estimates suggest that the elliptical instability could be responsible for the synchronisation of
the spins of hot Jupiters with their orbits if $P\lesssim 11.5$ d. In addition, this could be a very important mechanism contributing to orbital circularisation of initially eccentric planets, and could provide a partial explanation for the observed preponderance of circular orbits amongst the shortest period hot Jupiters out to $P\lesssim 4.5$ d. The elliptical instability is therefore worth studying in more detail, since if these estimates are found to in fact underestimate the dissipation, then this mechanism could explain the observed circularisation periods of hot Jupiters.

The above estimate may be incorrect. In particular, it is unclear whether the instability can lead to continual deposition of energy at an efficient rate once the saturated state of the resulting turbulence has been reached. 
To compare this estimate with our simulations, which we describe in the next few sections, we define the efficiency factor $\chi$, by
\begin{eqnarray}
D = \chi m_{1} (2R_{1})^{2} \gamma^{3} \epsilon^{3},
\end{eqnarray}
or $\chi \equiv D/\epsilon^3$ in the non-dimensional units adopted from \S 5 onwards. If $\chi$ is found to be small, 
the turbulence generated by this instability will produce a tidal $Q^{\prime}$ which is larger by a factor $\chi^{-1}$ than would be predicted by Eq.~\ref{ordermagq}. Given the potential importance of this mechanism, it is essential to study this problem in more detail and to quantify the efficiency of dissipation, $\chi$, and its dependence on $\epsilon, \Omega$ and $\nu$, by performing numerical simulations. This is what we shall attempt in the rest of this paper.

\section{Local model of a tidally deformed body}

Our model consists of a gaseous star or planet subject to an elliptical deformation resulting from the tidal perturbations of a companion body. We consider the primary to rotate with spin vector $\boldsymbol{\Omega}$, and the companion to orbit the primary with rotation vector $\boldsymbol{n}$, where these are assumed to be parallel (i.e. we neglect the problem of studying the instability when there is a spin--orbit misalignment). We consider a small patch located within the rotating primary body, which is sufficiently small that it can be considered homogeneous, with uniform density $\rho$ (henceforth taken to be unity, without loss of generality). In this patch, the flow in the primary body can be approximated as that of a cylindrical vortex, subject to a (in general, time-dependent) strain, which represents the equilibrium tide response in the rotating primary body. 

A local model has the advantage that it allows high-resolution numerical calculations of the nonlinear outcome of perturbations to this background flow, and is absent of the complications of boundary layers. We also neglect stratification, and therefore our calculations only strictly apply to a small patch of an adiabatically stratified convective region of a planet or star, where buoyancy forces are unimportant. Studying the elliptical instability in the presence of turbulent convection is also not attempted in this paper, so that we can study this instability in isolation. This model should allow us to study the properties of the turbulence generated through the elliptical instability, and to quantify the resulting dissipation rate, from which we can estimate its astrophysical importance.

We intend to study the instability of a background flow with elliptical streamlines, which has velocity field $\mathrm{A}_{n} \boldsymbol{x}$, in the frame centred on the primary body, and which is rotating with the orbital angular velocity $\boldsymbol{n}$, where
\begin{eqnarray}
\mathrm{A}_{n} = -\gamma \left(\begin{array}{ccc}
0 &  -(1+\epsilon) & 0 \\
1-\epsilon & 0 & 0 \\
0 & 0 & 0 \end{array}\right),
\end{eqnarray}
and $\epsilon < 1$.

We will work in the frame co-rotating with the spin of the primary body, with angular velocity $\boldsymbol{\Omega}=\Omega \mathbf{e}_{z}$, in which the strain is periodic in time. 
The base flow is then
\begin{eqnarray}
\boldsymbol{U}_{0} = \mathrm{A}\boldsymbol{x},
\end{eqnarray} 
with $\mathrm{A}$ being independent of the coordinates (but depends on time). The ``background vortex'' in this new frame takes the form of the oscillatory shear flow (e.g.~\citealt{Goodman1993})
\begin{eqnarray}
\label{timedepA}
\mathrm{A} = -\gamma \epsilon\left(\begin{array}{ccc}
s_{2\gamma t} &  c_{2\gamma t} & 0 \\
c_{2\gamma t} & -s_{2\gamma t} & 0 \\
0 & 0 & 0 \end{array}\right),
\end{eqnarray}
where $\cos 2\gamma t \equiv c_{2\gamma t}$ and  $\sin 2\gamma t \equiv s_{2\gamma t}$.
This represents the simplest response in the star to the tidal forcing due to a non-synchronous secondary on a circular coplanar orbit (these assumptions can be relaxed by adding additional components to the matrix A). We consider perturbations to this flow $\boldsymbol{u}$, 
such that the total velocity is $\boldsymbol{U}_{0}+\boldsymbol{u}$, and assume that $\boldsymbol{U}_{0}$ is perfectly maintained, so there is an infinite reservoir to drive the instability. 

Since our local patch is homogeneous, and we are interested in low frequency phenomena, we consider an incompressible three-dimensional rotating fluid, satisfying the Navier-Stokes equations for the evolution of perturbations to the base flow
\begin{eqnarray}
\label{NS1}
&& D \boldsymbol{u} + \boldsymbol{u}\cdot \nabla \boldsymbol{U}_{0} + 2\boldsymbol{\Omega}\times\boldsymbol{u} = -\nabla p + D_{\alpha}(\boldsymbol{u}), \\
\label{NS2}
&& \nabla \cdot \boldsymbol{u} = 0, \\
\label{NS3}
&& D \equiv \partial_{t}+ \boldsymbol{U}_{0}\cdot \nabla + \boldsymbol{u}\cdot \nabla,
\end{eqnarray}
subject to appropriate boundary conditions, where $\boldsymbol{u}$ is the velocity, and $p$ is the pressure. The viscous diffusion operator has been written as a generalised hyperviscosity $D_{\alpha}(\boldsymbol{u}) = (-1)^{\alpha+1}\nu_{\alpha}\nabla^{2\alpha}\boldsymbol{u}$, which reduces to the Navier-Stokes viscosity when $\alpha=1$, where the kinematic viscosity is $\nu_{1}$. We will also use $\alpha =4$, for which (with an appropriate choice of $\nu_{\alpha}$) dissipation is restricted to the smallest scales. 
Eqs.~\ref{NS1}--\ref{NS3} is the system of equations that we will solve numerically\footnote{
Note that our model can be considered to represent either a small patch located near to the centre of the body, or a patch that is located at a distance $\boldsymbol{x}_{c}(t)$ from the centre, where $\dot{\boldsymbol{x}}_{c}=\mathrm{A}\boldsymbol{x}_{c}$, so that the patch follows a streamline of the elliptical flow. To see this, transform into the frame centred at $\boldsymbol{x}_{c}(t)$, with coordinates $\boldsymbol{x}^{\prime}=\boldsymbol{x}-\boldsymbol{x}_{c}$. The time derivative in this frame transforms according to $\left(\partial_{t} \boldsymbol{u}\right)_{\boldsymbol{x}^{\prime}}=\left(\partial_{t} \boldsymbol{u}\right)_{\boldsymbol{x}} + \dot{\boldsymbol{x}}_{c}\cdot \nabla \boldsymbol{u}$, whereas the perturbed velocities in the two frames are related by $\boldsymbol{u}^{\prime}= \boldsymbol{u}$. Therefore, it can be seen that the equations are identical, but with primed quantities replacing those in the original frame. This is a consequence of the assumption that $\boldsymbol{U}_{0}$ is linear in the coordinates.}.

We introduce a non-dimensionalisation in which our unit of length is the dimension of the box $L$, and the time unit is $\gamma^{-1}$. Together with the values of $\Omega$ and $\epsilon$, this leaves one dimensionless quantity, the Reynolds number $\mathrm{Re} = \gamma L^{2}/\nu$ (strictly only defined as such for $\alpha=1$). We desire to make $\mathrm{Re}\gg1$ to minimise the effects of viscosity, but choose a value of $\nu_{\alpha}$ sufficiently large to properly resolve the flow. One further relevant dimensionless quantity is the Rossby number $\mathrm{Ro}=\frac{\epsilon \gamma}{\Omega}$, where $\mathrm{Ro} \ll1$, which tells us that the flow is likely to be strongly rotationally constrained for weak ellipticities.

\section{Spectral decomposition into shearing waves}

Our numerical method for simulating a small patch of an elliptical vortex exploits ideas similar to those involved in the construction of the shearing box in an accretion disc (e.g.~\citealt{GoldreichLyndenBell1965}; \citealt{Hawley1995}), that trace their origin to \cite{Kelvin1880}. We expand the perturbation into a set of shearing waves, which 
are plane waves with evolving wavevectors $\boldsymbol{k}(t)$:
\begin{eqnarray}
\left[\boldsymbol{u}(\boldsymbol{x},t),p(\boldsymbol{x},t)\right]= \mathrm{Re}\left\{\left[\hat{\boldsymbol{u}}(t),\hat{p}(t)\right] e^{i \boldsymbol{k}(t)\cdot \boldsymbol{x}}E_{\nu}(t)\right\}.
\end{eqnarray}
These form a complete basis if we adopt periodic boundary conditions in the flow comoving with $\boldsymbol{U}_{0}$.
We have introduced the viscous decay factor $E_{\nu}(t) = \exp ((-1)^{\alpha}\nu_{\alpha}\int^{t}k^{2\alpha}(t^{\prime})dt^{\prime} )$, which fully captures the effects of (hyper)viscosity and eliminates the (hyper)viscous terms from Eq.~\ref{NS1},
after spatial Fourier transforms have been performed. These shearing waves evolve according to the ordinary differential equations
\begin{eqnarray}
\partial_{t}\hat{\boldsymbol{u}} + \mathrm{A}\hat{\boldsymbol{u}} + 2\boldsymbol{\Omega}\times\hat{\boldsymbol{u}} + i\boldsymbol{k}\hat{p} = -\mathbf{\widehat{\boldsymbol{u}\cdot\nabla\boldsymbol{u}}},
\end{eqnarray}
\begin{eqnarray}
\label{kvol1}
\dot{\boldsymbol{k}} + \mathrm{A}^{T}\boldsymbol{k} = 0,
\end{eqnarray}
where T represents the transpose operator, together with the incompressibility constraint, 
\begin{eqnarray}
\boldsymbol{k} \cdot \hat{\boldsymbol{u}} = 0.
\end{eqnarray}
The last equation tells us that a single wave is an exact solution of Eqs.~\ref{NS1}--\ref{NS3} (i.e.~$\mathbf{\widehat{\mathbf{u}\cdot\nabla\mathbf{u}}}=0$). However, this is not true of the more realistic case, involving multiple shearing waves, because they couple nonlinearly, and can therefore transfer energy between different components.

\begin{figure}
  \begin{center}
    \subfigure{\includegraphics[trim=0cm 0cm 0cm 0cm, clip=true,width=0.5\textwidth]{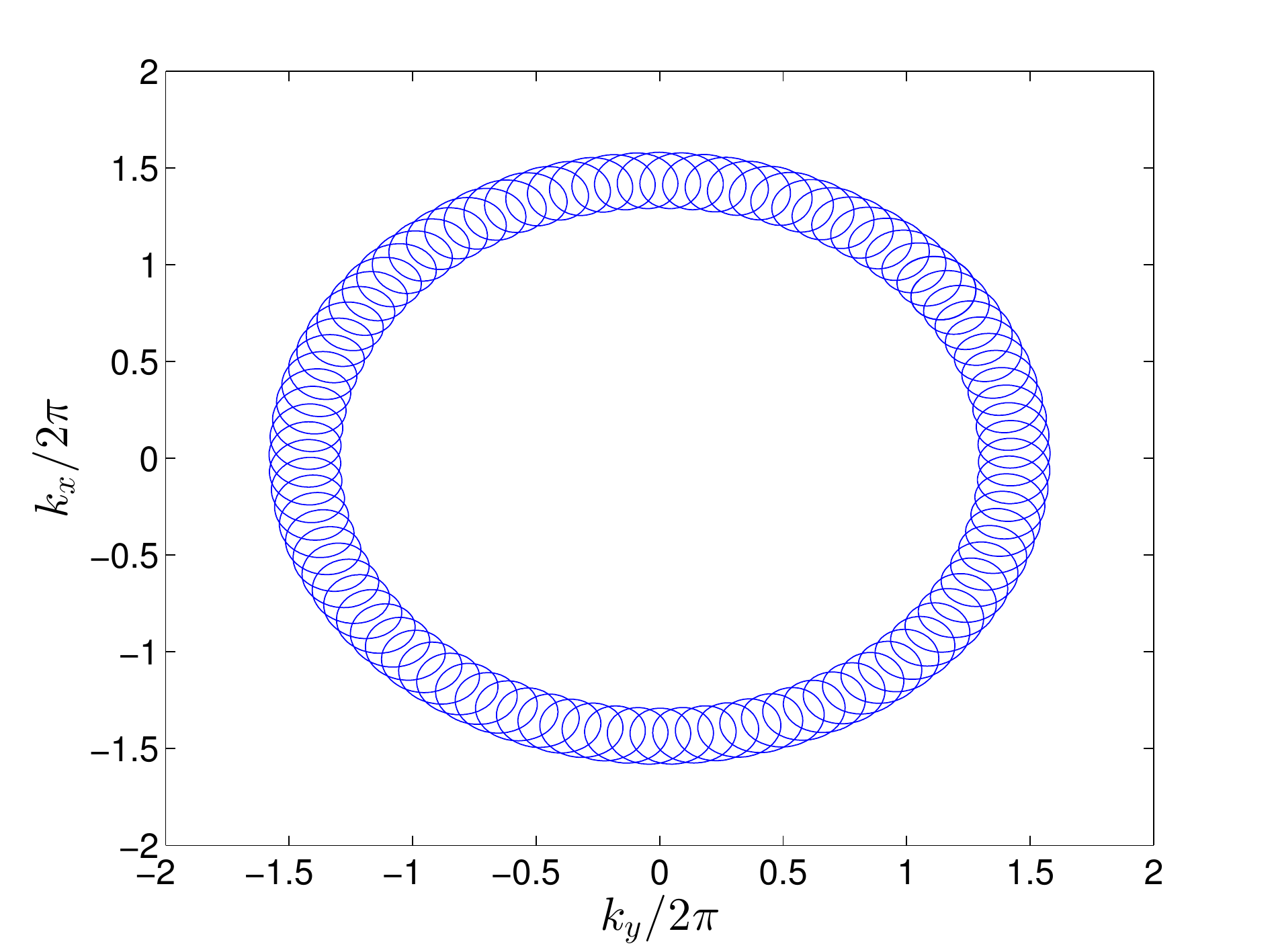} }
    \end{center}
      \caption{Illustration of the temporal evolution of the wavevector on the ($k_{x},k_{y}$)-plane according to Eq.~\ref{kvol1}, for an initial wave with $k_{x}(0)=k_{y}(0)=2\pi$, on top of a background vortex with $\epsilon=0.2$. The evolution of wavevector is plotted from $t=0$ until $t=100\pi$. There is a stretching at the rate 2 and a rotation at the rate $\epsilon^2/2$.}
  \label{ShearingBox}
\end{figure}

The wavevectors are obtained by solving  Eqs.~\ref{kvol1} \& \ref{timedepA}
(see Figure \ref{ShearingBox}). The
 solution involves two different processes: a periodic stretching/shortening with period $\pi$, together with a slow rotation of the wavevector with period $4\pi/\epsilon^2$. The former is responsible for parametrically exciting inertial waves. 
 
 Evolving the wavevector in time allows us to use a standard Cartesian spectral code using 
Fourier decomposition for this problem, since periodic boundary conditions are automatically satisfied by these shearing waves.

\subsection{Energy budget}
To help with analysing the flow, we construct a mean energy equation by taking the scalar product of Eq.~\ref{NS1} with $\mathbf{u}$, and integrating over the box of (initial) size $L$
 along each horizontal dimension and $L_{z} = \beta L$ along the vertical (where $\beta$ is the aspect ratio), leading to
\begin{eqnarray}
\partial_{t}K = I-D.
\end{eqnarray}
We define the averaging operation $\langle \chi \rangle =  \frac{1}{\beta L^{3}}\int_{V} \chi dV$, the mean kinetic energy to be
\begin{eqnarray}
K &=& \frac{1}{2} \langle|\mathbf{u}|^{2}\rangle,
\end{eqnarray}
and the mean dissipation rate to be,
\begin{eqnarray}
D &=& -\nu(-1)^{\alpha+1}\langle\mathbf{u}\cdot \nabla^{2\alpha} \mathbf{u}\rangle.
\end{eqnarray}
We also define
\begin{eqnarray}
I = -\langle \mathbf{u}\mathrm{A}\mathbf{u}\rangle,
\end{eqnarray}
which represents the energy injection into the flow from the background, or more generally, the interaction between the flow and the background, since energy can, in principle, be transferred in either direction. 
If $I$ is positive, the perturbation kinetic energy will grow at the expense of the background vortex, otherwise the perturbation will decay by transferring its energy to the background vortex. For a single wave, this can either be positive or negative, depending on the initial wave phase.

The total kinetic energy of the background vortex does not secularly change from a value $K_0=\beta\frac{L^{5}}{12}\left(\Omega^{2}+(\gamma\epsilon)^{2}\right)$), in our simulations, since we evolve the wavevectors as if the background flow is perfectly maintained. In an astrophysical system, the elliptical flow would be maintained by the companion, and dissipation of energy would come at the expense of the spin of the body or the companion's orbit.

In rotationally dominated flows, anisotropy is introduced between the vertical and horizontal directions, and it is useful to decompose the velocity field into suitable components to describe this anisotropy. For reasons that will become clear, we decompose the flow into 2D components with $k_{z}=0$ (``vortices") and 3D components with $k_{z}\ne 0$ (``waves"). The corresponding mean energies for each of these components are defined by
\begin{eqnarray}
\label{rotmean1}
E_{3D} &=& \frac{1}{2}\sum_{\boldsymbol{k}| k_{z}\ne 0} |\hat{\boldsymbol{u}}(\boldsymbol{k},t) |^{2}, \\
E_{2D} &=& \frac{1}{2}{\sum_{\boldsymbol{k}| k_{z}= 0,k\ne 0}} |\hat{\boldsymbol{u}}(\boldsymbol{k},t) |^{2}, 
\label{rotmean2}
\end{eqnarray}
such that $K=E_{2D}+E_{3D}$. The evolutionary equations for each of these quantities read
\begin{eqnarray}
\partial_{t} E_{3D} &=& -T_{23}+I_{3D}- D_{3D}, \\
\partial_{t} E_{2D} &=& T_{23}+I_{2D}- D_{2D},
\end{eqnarray}
where $D_{i}$ are the corresponding mean dissipation rates for each component. We define the energy flux
\begin{eqnarray}
T_{23} &=& -\langle\boldsymbol{u}_{2D}\cdot\left(\boldsymbol{u}_{3D}\cdot\nabla\right)\boldsymbol{u}_{3D}\rangle,
\end{eqnarray}
to represent the transfer of energy due to nonlinear interactions between the waves and the vortices. We also define
\begin{eqnarray}
I_{3D} &=&  \langle\boldsymbol{u}_{3D}A\boldsymbol{u}_{3D}\rangle,
\end{eqnarray}
and similarly for $I_{2D}$, which represent the energy injection into the waves and the vortices from the background vortex. These equations allow us to track the temporal evolution of the flow anisotropy, and help to quantify the relevant processes involved.

\section{Numerical method}

Our method of solution is to use the Cartesian pseudo-spectral code SNOOPY (\citealt{Lesur2005}; \citealt{Lesur2007}). The code uses a 3rd-order Runge-Kutta method for time evolution, with the diffusive terms integrated using an integrating factor. The nonlinear terms are dealiased by applying the $3/2$ rule. We have modified the code so that it solves Eqs.~\ref{NS1}--\ref{NS3} by expanding the flow in terms of shearing waves  that satisfy Eq.~\ref{kvol1}.
Evolving the wavevector in this way is different from the shearing box representation of an accretion disc, in practice, by the fact that remapping (e.g.~\citealt{Lithwick})
is not required because $|\boldsymbol{k}(t)|$ does not increase without bound. Using a local model of this sort allows higher resolution of the turbulence than in a global simulation.

\subsection{Tests}

\begin{figure}
  \begin{center}
          \subfigure{\includegraphics[trim=0cm 0cm 0cm 0cm, clip=true,width=0.48\textwidth]{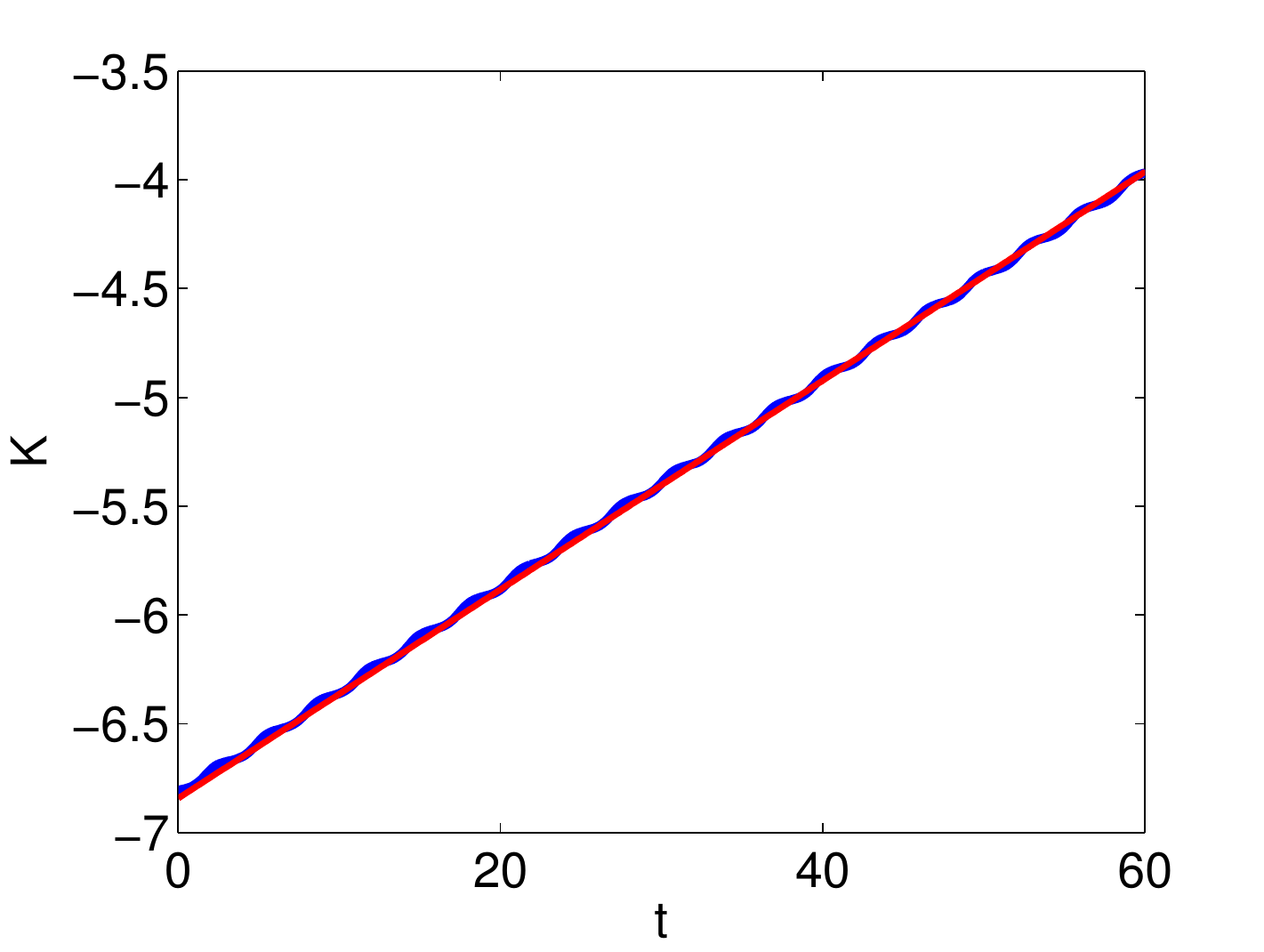} }
    \end{center}
  \caption{Illustrative test of the code at capturing the linear growth of the elliptical instability when $\gamma=\Omega=1$ and $\epsilon=0.1$, showing the computed kinetic energy (blue) compared against a linear profile with slope $9/8 \epsilon$ (red), corresponding with the maximum growth rate of the instability. Nonlinear and viscous terms were neglected, and a wave with integer wavenumbers $L k_{x}/2\pi=L k_{y}/2\pi=11,L k_{z}/2\pi=9$ (where the initial box length is $L$)
  was initialised at $t=0$ using $64^3$ modes. This mode has $k_{z}/k \approx 0.5$.}
  \label{lintest}
\end{figure}

Our first test was to ensure that our modifications to the code could correctly capture the evolution of a single shearing wave in the linear approximation, for which analytical solutions are available e.g. \cite{Kerswell2002} -- to do this we simply set the nonlinear terms to zero, so that this wave would not couple with noise level waves and lead to instability. We then tested the growth rates of the linear elliptical instability for several initial plane waves, which were accurately reproduced by the simulations -- an example is illustrated in Fig.~\ref{lintest}. We verified that the elliptical instability does not occur in the range $\Omega \in (-0.5,0.5)$. We also checked that a plane inertial wave (with $\epsilon=0$) is unstable to secondary parametric instabilities, whose growth rates have the correct scaling with amplitude and wavenumber as expected (e.g.~\citealt{LifshitzFabijonas1996}). Finally, we checked that the energy equation is accurately satisfied in our solution, and that the relative energy error is small. This should convince us that the code is able to accurately resolve the flows described below.

\section{Nonlinear evolution of the elliptical instability: Results}

In this section we first describe the results of a fiducial simulation in a unit box ($\beta=1$), with $\epsilon=0.1$, using a resolution of $256^3$ and standard viscosity with $\alpha=1$, $\nu_{1}=10^{-5}$. We initialise large scale noise with wavenumbers in the range $1 \leq k/(2\pi) \leq 12$, with small random amplitudes ($\sim 0.01$) and random phases. The results are found to be relatively insensitive to variations in the wavenumber range and amplitudes of the initial conditions. For simplicity we first consider the case in which $n=0$ and $\Omega=1$. This situation is clearly unphysical, since it would correspond with a stationary secondary body, however it is the simplest in which to begin our investigation -- in addition, having $n=0$ as opposed to $n\ne 0$ does not qualitatively affect our results. The effects of a deformation that is rotating in the inertial frame, resulting from the orbital motion of the secondary body, will be described in section \S~\ref{rotatingstrain} below.

\subsection{Fiducial simulation with Navier-Stokes viscosity}
\label{fid1}

\begin{figure}
  \begin{center}
      \subfigure{\includegraphics[trim=0cm 0cm 0cm 0cm, clip=true,width=0.4\textwidth]{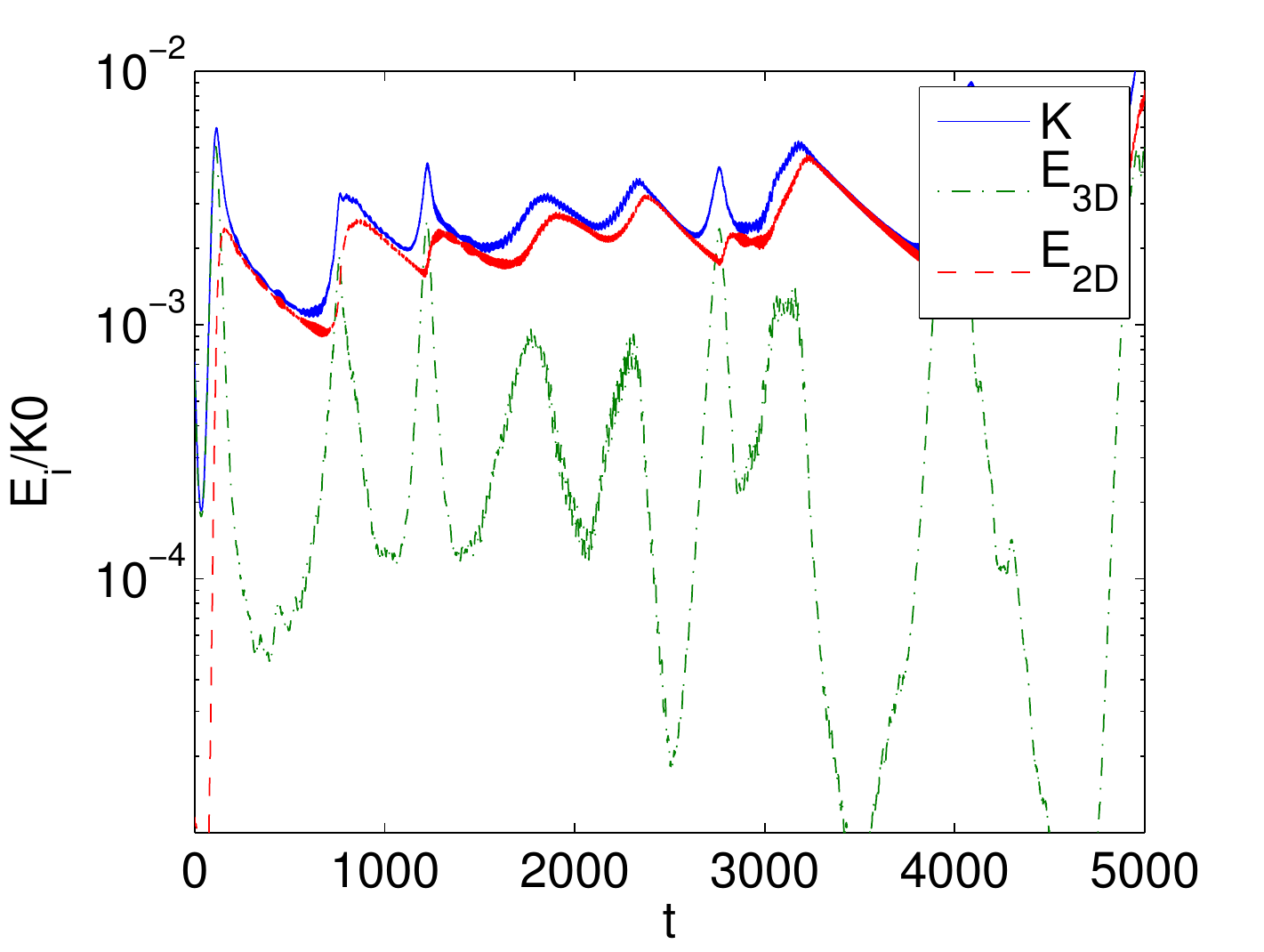} }
          \subfigure{\includegraphics[trim=0cm 0cm 0cm 0cm, clip=true,width=0.4\textwidth]{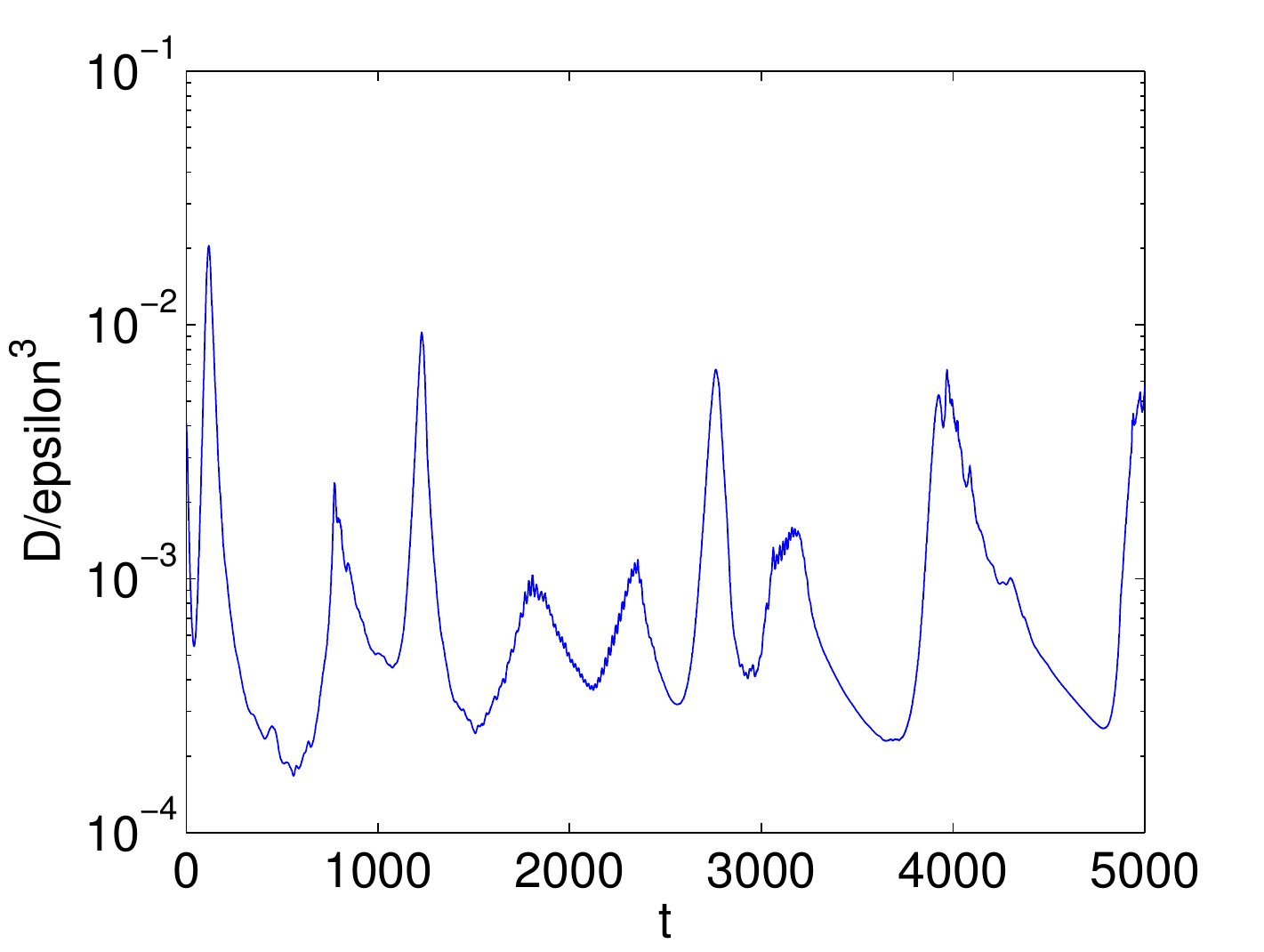} } 
          \subfigure{\includegraphics[trim=0cm 0cm 0cm 0cm, clip=true,width=0.4\textwidth]{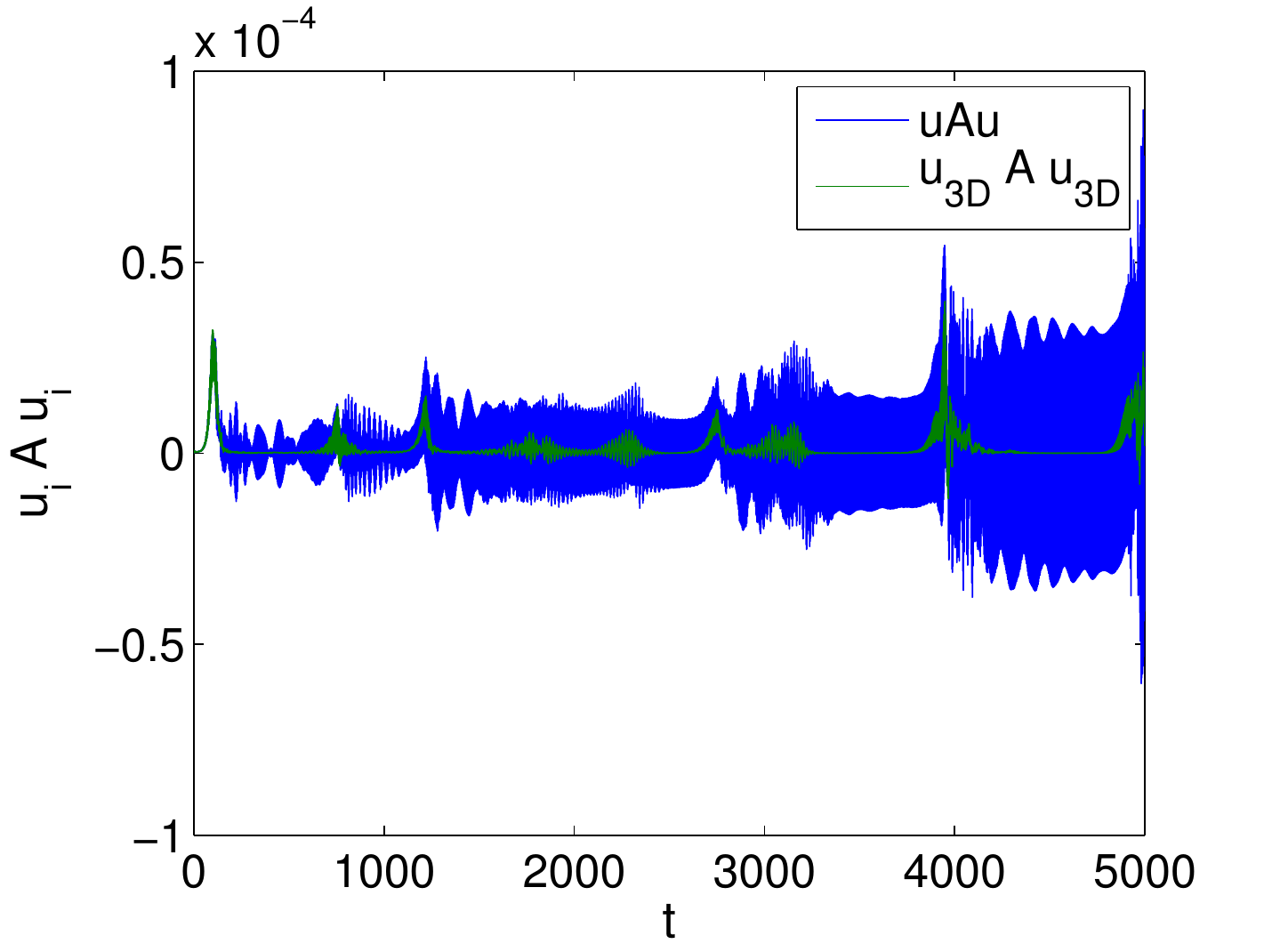} } 
             \subfigure{\includegraphics[trim=0cm 0cm 0cm 0cm, clip=true,width=0.4\textwidth]{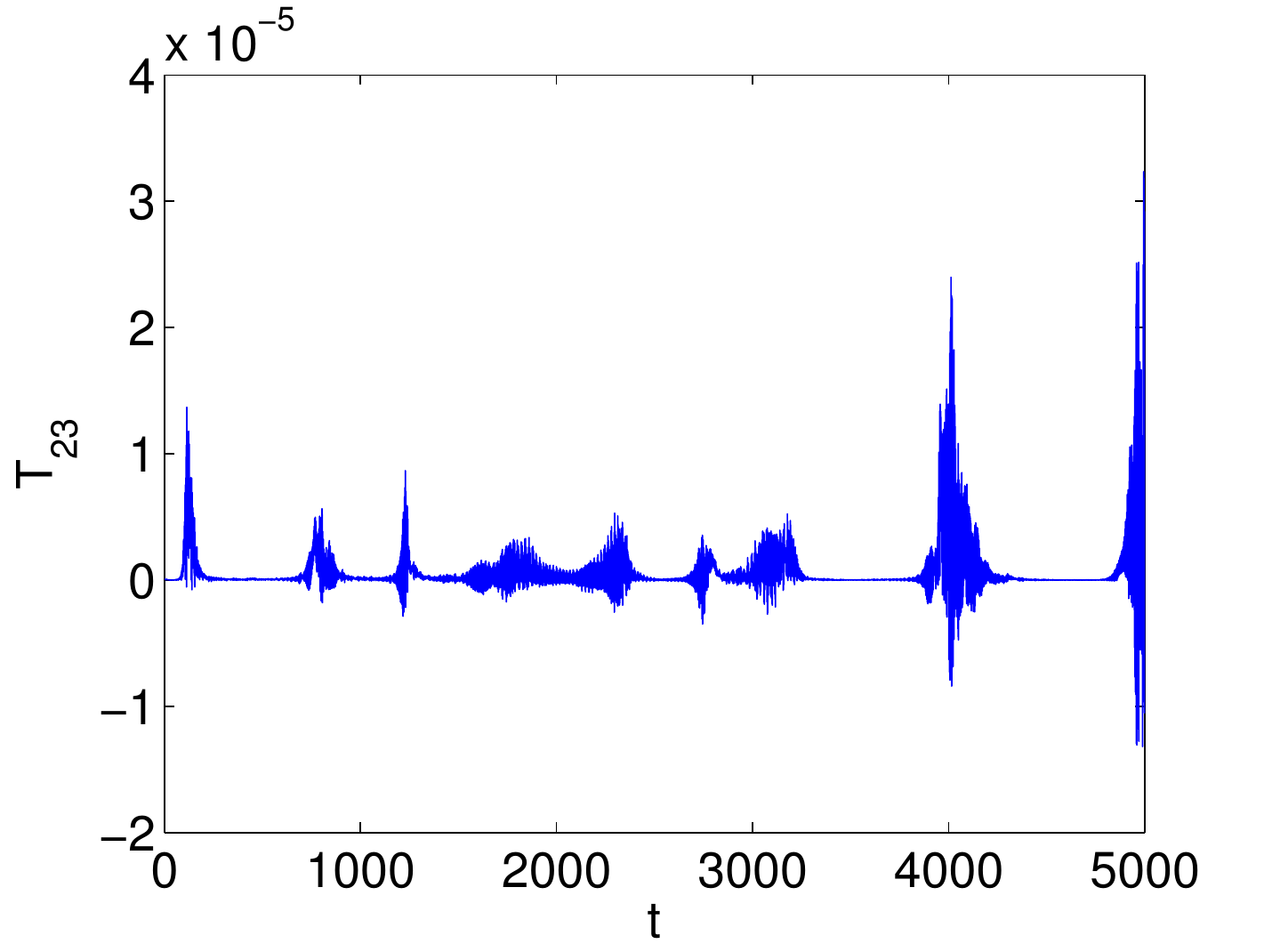} } 
    \end{center}
  \caption{
  Temporal evolution of our fiducial simulation with $n=0$, $\epsilon=0.1$, $\alpha=1$ and $\nu=10^{-5}$ in a unit box ($\beta=1$). 
  This shows kinetic energy, normalised to the background vortex (top), dissipation normalised to $\epsilon^3$ (top middle), energy injection into the total flow and the waves (bottom middle), and finally, energy transfers from the waves to the vortices (bottom). }
  \label{tvol}
\end{figure}

In the initial stages, the periodic strain excites modes with the correct orientations and phases to be driven by the background, having frequencies comparable with $\gamma$ and phases corresponding with the stretching directions of the background vortex.
During this phase, the flow is composed of a superposition of exponentially growing inertial waves, which only weakly interact. The evolution of the kinetic energy of the flow, normalised to that of the background vortex, is presented in the top panel of Fig.~\ref{tvol}. 

\begin{figure}
  \begin{center}
   \subfigure{\includegraphics[trim=1cm 17cm 1cm 1cm, clip=true,width=0.48\textwidth]{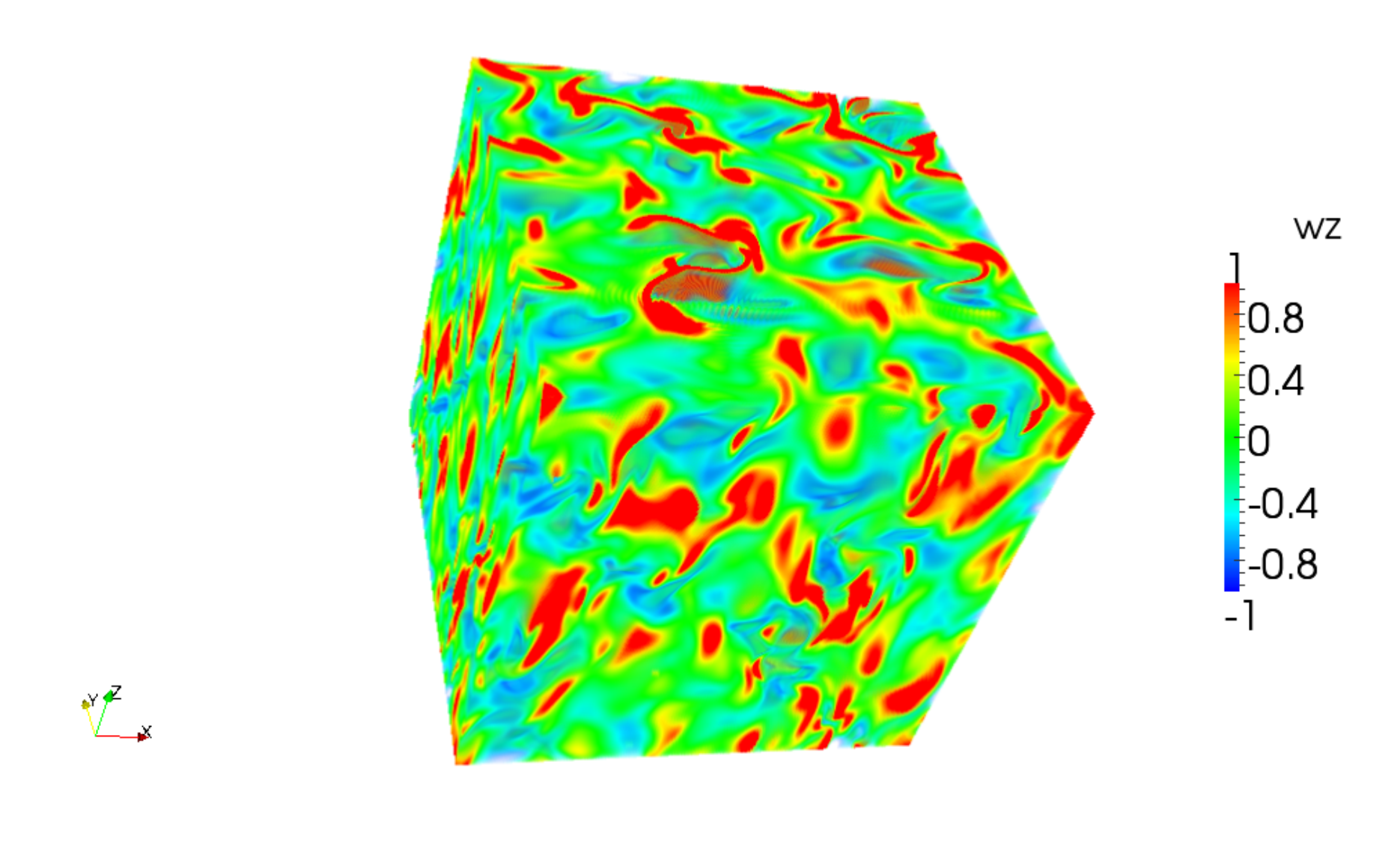} } 
      \subfigure{\includegraphics[trim=1cm 17cm 1cm 1cm, clip=true,width=0.48\textwidth]{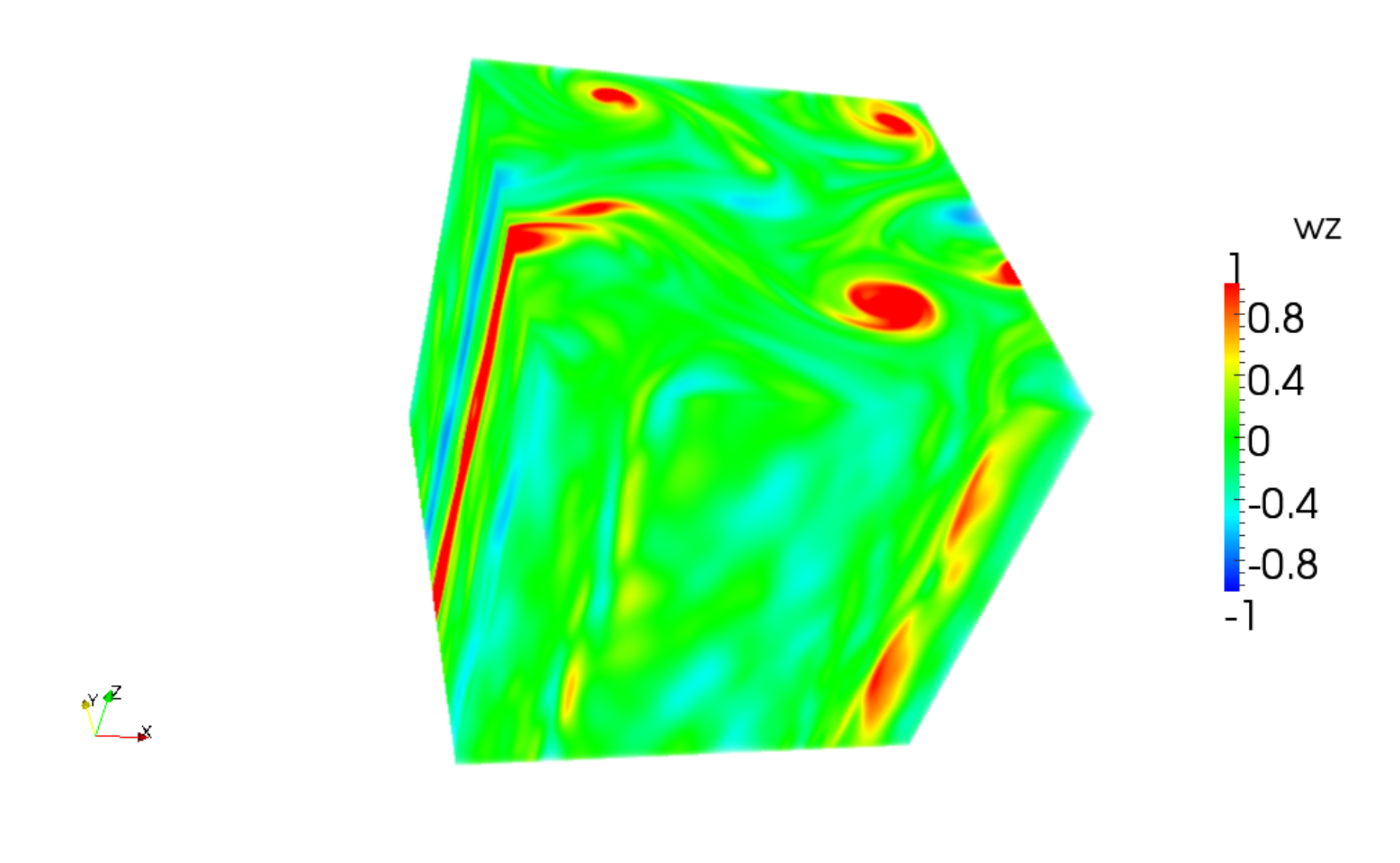} } 
          \subfigure{\includegraphics[trim=1cm 17cm 1cm 1cm, clip=true,width=0.48\textwidth]{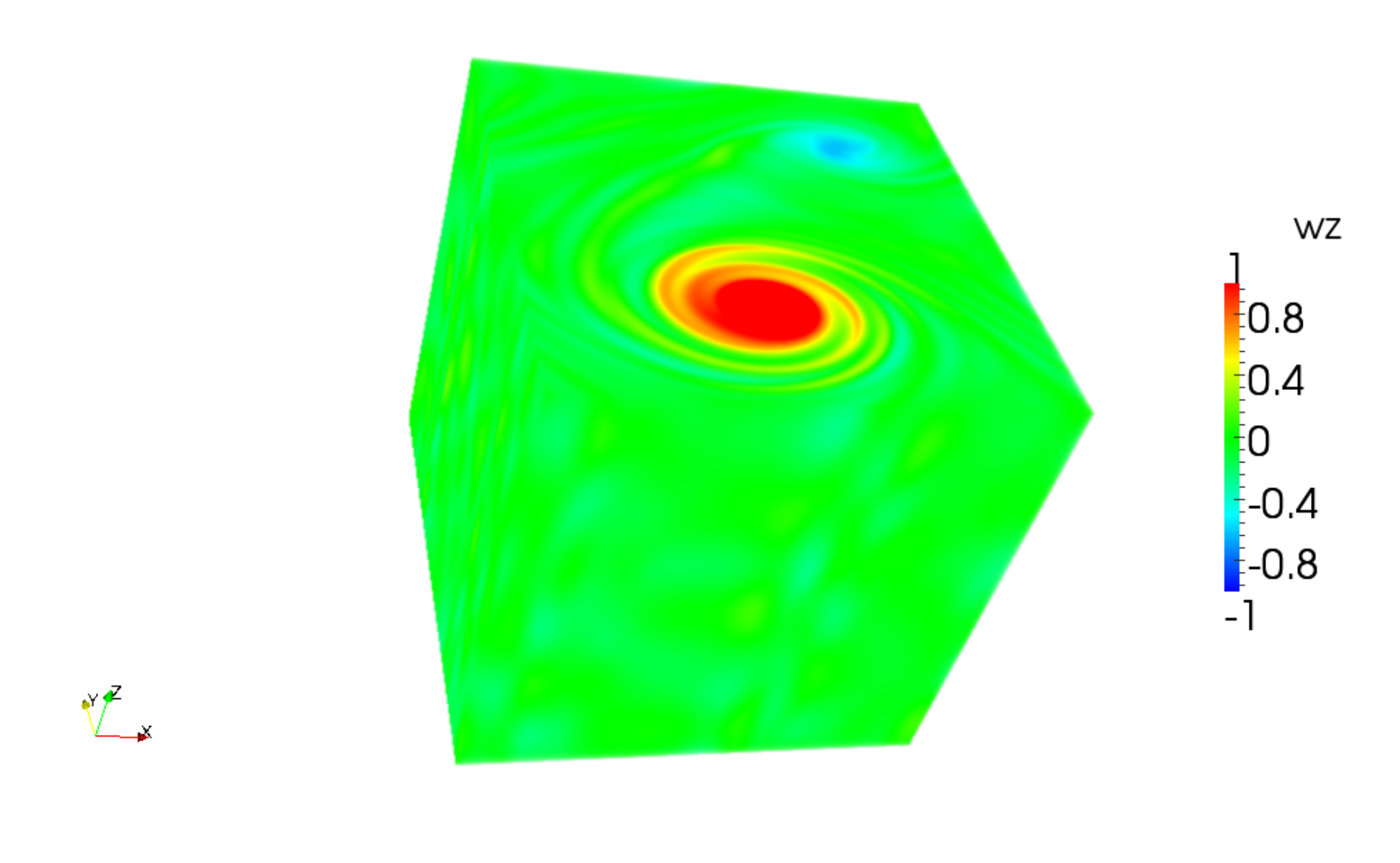} } 
    \end{center}
  \caption{Volume renderings of the vertical vorticity $\omega_{z}$ in our fiducial simulation with $\alpha=1$ and $\nu=10^{-5}$. These snapshots are at $t=100,200,4200$.}
  \label{vol}
\end{figure}

Once the excited modes attain sufficient amplitude (which occurs here when $t\sim 100$), nonlinear effects become important. Secondary instabilities set in and rapidly break the growing waves, causing a prompt transition to turbulence. An example of the flow field in this stage is given in the first panel of Fig.~\ref{vol}. During this transient turbulent phase, the flow becomes dominated by columnar vortices aligned with the rotation axis (``Proudman-Taylor columns"), as is illustrated in the second panel of Fig.~\ref{vol}. Such structures are naturally produced by nonlinear interactions in rapidly rotating turbulence with Ro $\ll 1$, which is always true for the bulk of our flow when $\epsilon \ll 1$. 
This process occurs over several rotation periods, and creates an approximately two-dimensional, two-component flow, with the root mean square values of $|u_{x}|\sim |u_{y}| \sim \epsilon$ and $|u_{z}|\ll |u_{x,y}|$. This tendency towards two-dimensionalisation is quantified in the bottom panel of Fig.~\ref{tvol}, where we have plotted the total nonlinear energy transfers from the waves to the vortices. The quantity $T_{23}$ is positive shortly after $t\sim 100$, indicating that the waves are feeding the vortices.

The subsequent evolution is dominated by the interactions between these strong columnar vortices. The dissipation rate is drastically reduced, and the system evolves much more slowly compared with the transient turbulent phase. In the top panel of Fig.~\ref{tvol} we have decomposed the energy into the components defined in Eqs.~\ref{rotmean1}--\ref{rotmean2}. From $t\sim 100-500$, the kinetic energy is dominated by strong long-lived columnar vortices, since $K\sim E_{2D}$, with the energy in the waves contributing only a tiny fraction of the total energy. Most importantly, the energy injection rate into the waves ($I_{3D}$) is significantly reduced in the presence of strong vortices, which shows that the primary elliptical instability is inhibited in their presence. This can be seen in the third panel of Fig.~\ref{tvol}. 
Note that $I$ is rapidly oscillating, with fluctuations much larger than the mean, in contrast to $I_{3D}$, implying that this behaviour primarily represents the injection of energy into/from the vortices, which approximately cancels, on average. Note that vortices are not directly forced, since the elliptical instability excites inertial waves through a parametric resonance (these must therefore have nonzero frequencies). Vortices must result from the nonlinear interactions between inertial waves.

The vortices interact and merge on a timescale primarily controlled by the viscosity (see e.g.~\citealt{Meunier2005}). Once like-signed vortices have merged the system usually consists of a pair of vortices, as seen in the bottom panel of Fig.~\ref{vol}, with one strong cyclonic vortex and a somewhat weaker anticyclonic vortex. The peak vorticity amplitudes are $O(\gamma)$ (or larger), however, the mean magnitude of vorticity in the flow is smaller, and is $O(\epsilon)$. The total kinetic energy contained in the vortices is much less than that of the background flow $\boldsymbol{U}_{0}$. This phase persists for a significant fraction of the viscous decay time of the dominant vortices.

Once the vortices have viscously damped to below a particular energy 
(corresponding with $K/K_{0}\sim 10^{-3}$ in this example), $I_{3D}$ increases, and the waves are again efficiently driven at $t\sim 700$. These waves exponentially grow and subsequently transfer their energy into the vortices, which again prevent excitation of the waves. This cyclic behaviour is observed to recur throughout the simulation, on a timescale related to the viscous decay timescale of the dominant cyclonic vortices. Note that the global viscous timescale is $\nu^{-1}=10^{5}$, whereas that for the vortices is somewhat smaller and is $\sim 10^{3}$. The dissipation peaks when the waves are efficiently driven by the background, but is strongly reduced when the vortices are dominant in the flow. It is interesting to note that similar ``bursting" behaviour, in which the turbulent intensity quasi-periodically grows and then decays, has also been observed in previous laboratory experiments and numerical simulations of the elliptical instability (e.g.~\citealt{Malkus1989,Lacaze2004,Cebron2012}).

It appears that the presence of strong vortices, which dominate the flow-field, prevents the efficient excitation of waves. This is presumably a result of the specific phase and frequency dependence of the instability, since only those waves with frequencies approximately $\gamma$ and particular phases are excited. The presence of vortices is expected to detune the frequencies, and more importantly, the phases of the modes, producing a stabilising effect, which prevents the waves from being coherently driven. The wave will not spend sufficient time with the correct phase to allow sustained energy injection when it is perturbed by strong vortices. 

To summarise the evolution of our fiducial simulation, the elliptical instability excites inertial waves, which become unstable at finite-amplitude to secondary instabilities, resulting in transient turbulence. Strong columnar vortices are produced by nonlinear interactions between inertial waves, and these dominate the resulting flow field. Once vortices have formed, they merge, eventually producing a pair of vortices of opposite sign. The elliptical instability saturates through the formation of strong columnar vortices, whose presence effectively suppresses the driving mechanism. This results in a significantly weaker dissipation than we might expect if the wave driving is sustained. In the next section we will briefly present the differences observed with a similar simulation using hyperviscosity, instead of standard viscosity, to illustrate the importance of viscous damping of the large scale vortices, before presenting the results from varying the relevant parameters of our problem.

\subsection{Fiducial simulation with hyperviscosity}
\label{hypervis}

\begin{figure}
  \begin{center}
      \subfigure{\includegraphics[trim=0cm 0cm 0cm 0cm, clip=true,width=0.4\textwidth]{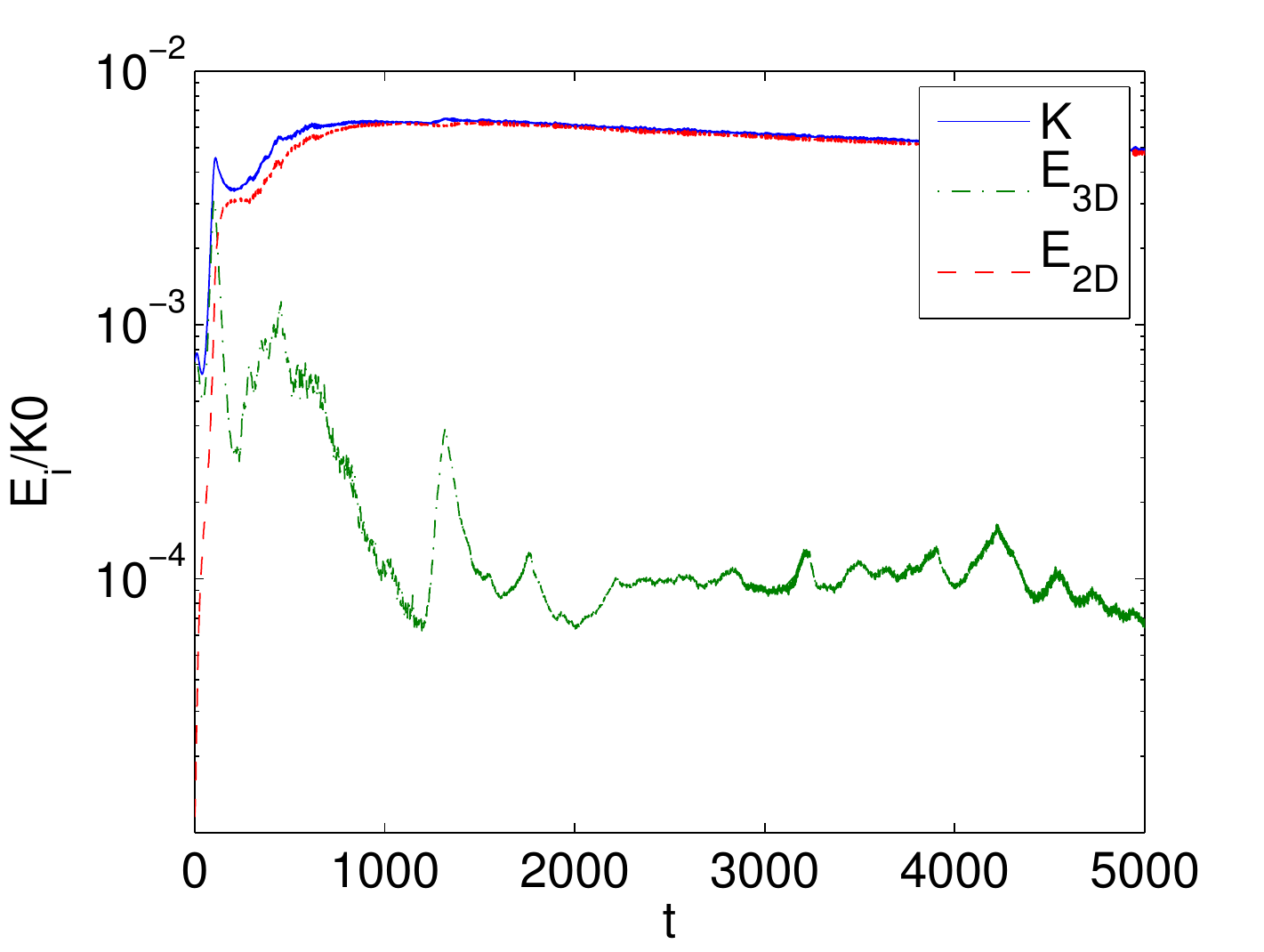} }
          \subfigure{\includegraphics[trim=0cm 0cm 0cm 0cm, clip=true,width=0.4\textwidth]{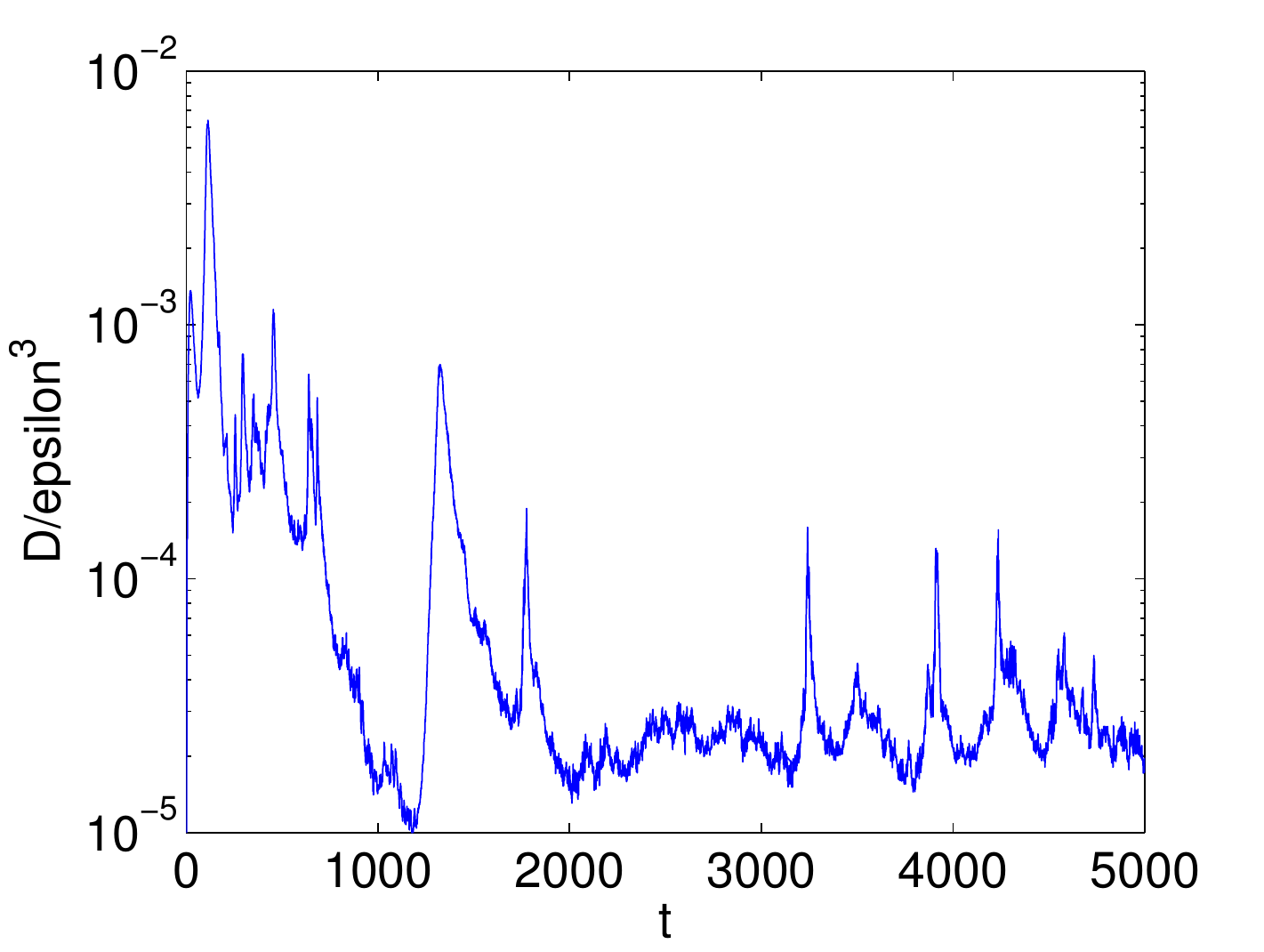} } 
          \subfigure{\includegraphics[trim=0cm 0cm 0cm 0cm, clip=true,width=0.4\textwidth]{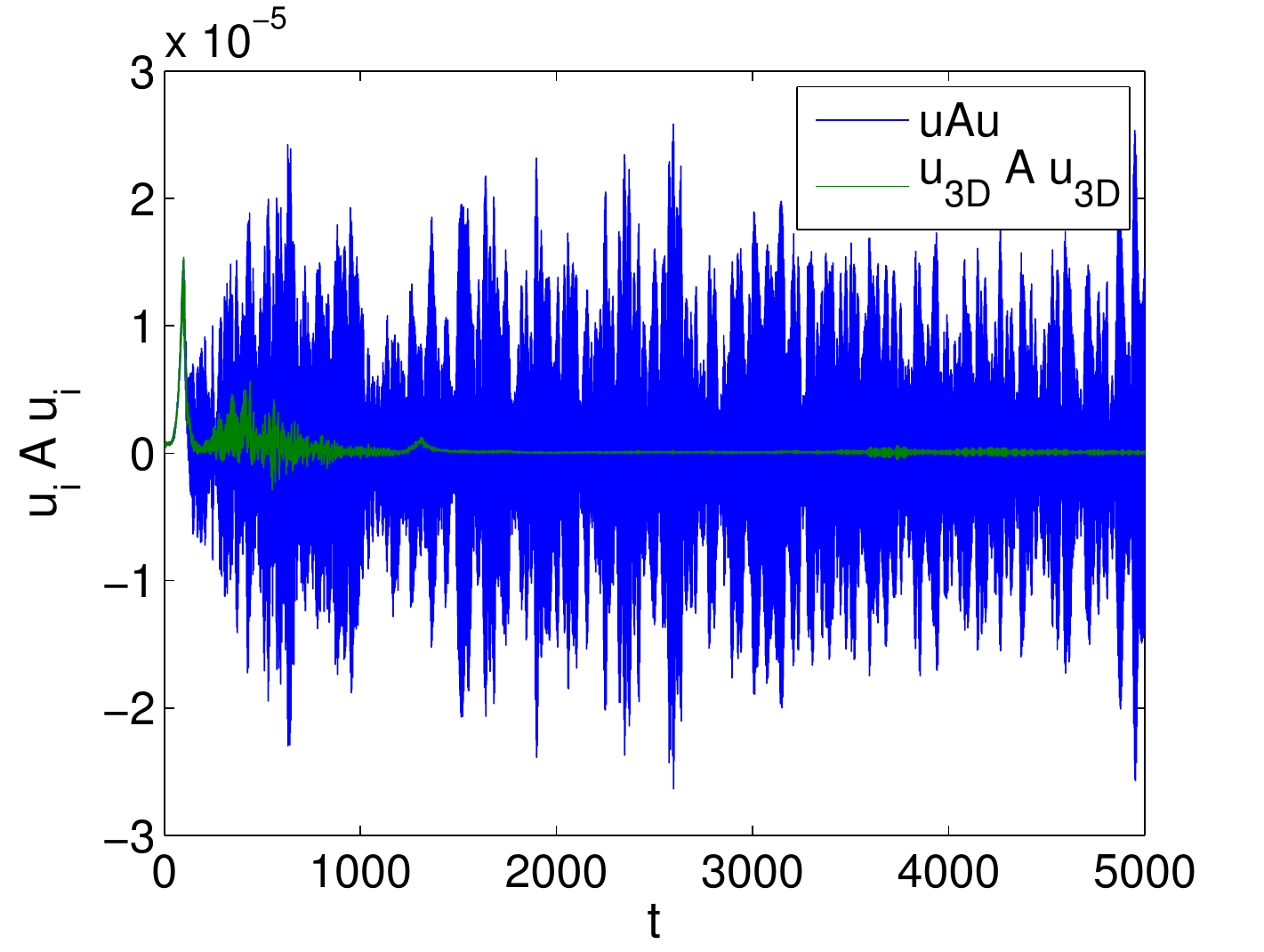} } 
            \subfigure{\includegraphics[trim=0cm 0cm 0cm 0cm, clip=true,width=0.4\textwidth]{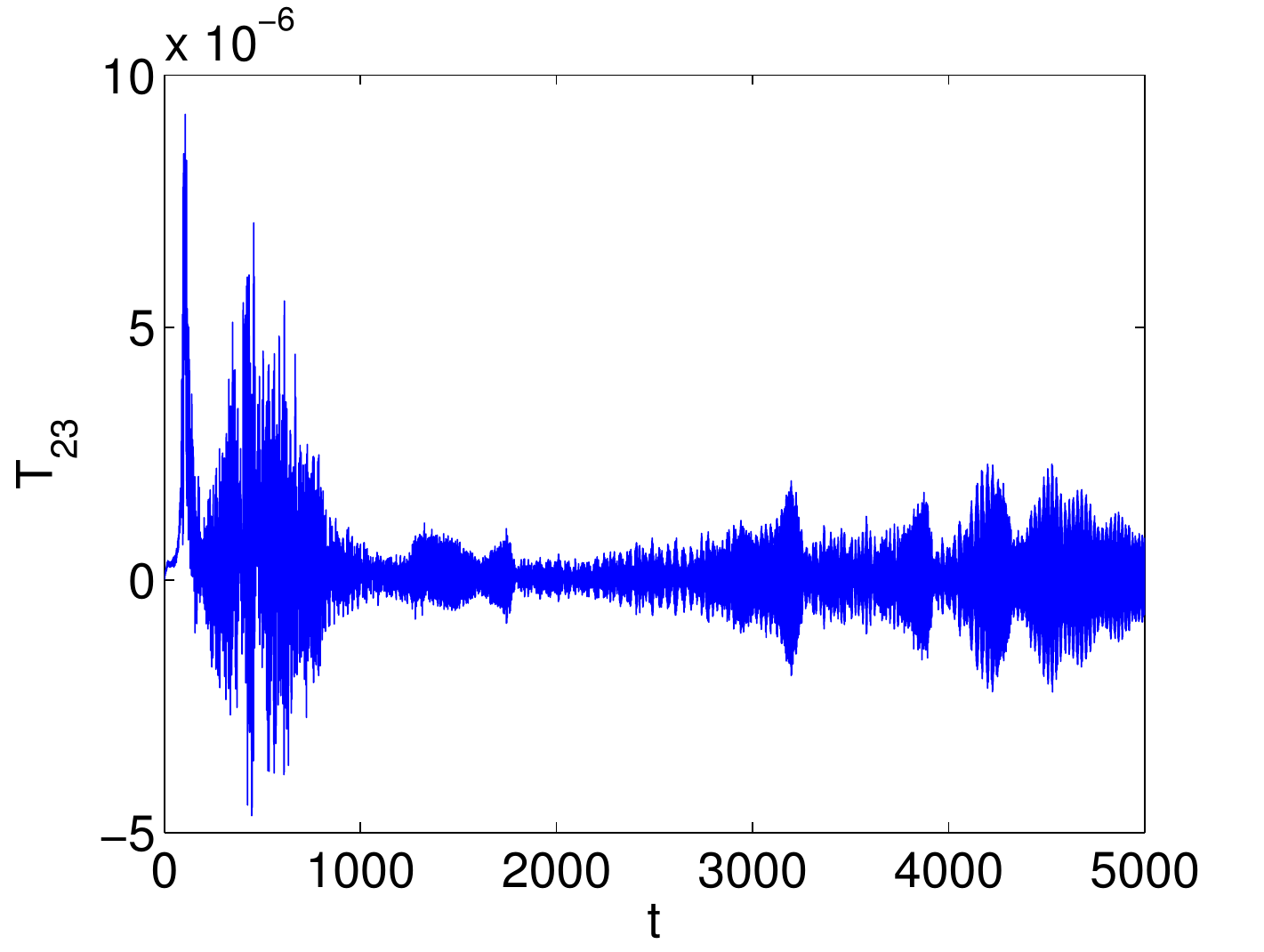} } 
    \end{center}
  \caption{Same as Fig.~\ref{tvol} except for a simulation with $n=0$, $\epsilon=0.1$, $\alpha=4$ and $\nu_{4}=10^{-20}$.} 
  \label{tvol2}
\end{figure}

\begin{figure}
  \begin{center}
         \subfigure{\includegraphics[trim=1cm 14cm 1cm 1cm, clip=true,width=0.48\textwidth]{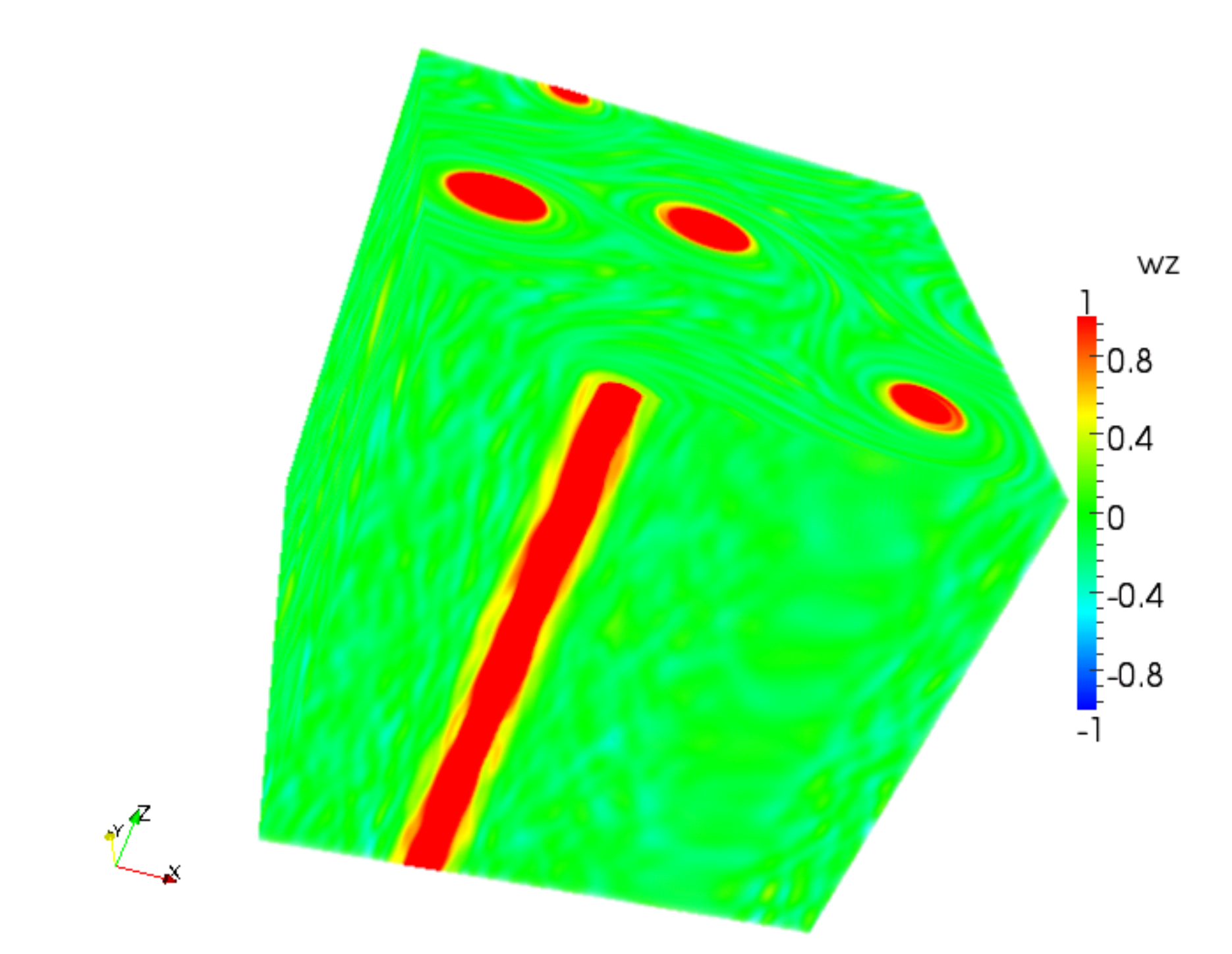} } 
    \end{center}
  \caption{Volume rendering of the vertical vorticity in our fiducial simulation with hyperviscosity at $t=4600$ with $\epsilon=0.1$, $\alpha=4$ and $\nu_{4}=10^{-20}$.}
  \label{vol2}
\end{figure}

The simulation above used standard viscosity, with $\alpha=1$. In this section we will present a simulation using $\alpha=4$ and $\nu_{4}=10^{-20}$, which is otherwise the same. The reason for adopting this form of the dissipative operator is to eliminate any effects that occur on the viscous timescale of the large-scale vortices in the previous simulation. Hyperviscosity restricts the dissipation to the smallest scales. In this simulation, the global viscous timescale is $10^{20}$, whereas that for the vortices is $\sim 10^{12}$, both of which are much longer than the duration of the simulation.

It can be seen in Fig.~\ref{tvol2} that the initial evolution is very similar using hyperviscosity. Once columnar vortices have formed, they merge at a slower rate with hyperviscosity, so it takes longer for the final state involving a pair of vortices to be produced. In fact, by $t\sim5000$, the flow is still composed of 4 cyclonic vortices instead of the single cyclonic vortex produced much earlier in the $\alpha=1$ simulation, e.g.~see Fig.~\ref{vol2}. The peaks in the dissipation correspond with discrete vortex merger events, being much weaker in between mergers, where the vortices make a negligible contribution to the dissipation.
 
Cyclic behaviour is not observed, presumably because this requires evolution for a significant fraction of the viscous timescale for the vortices, which is unattainably large when hyperviscosity is used. This means that the dissipation remains weak for much longer, and is generally much weaker than the case with $\alpha=1$, in which the dissipation of the vortices is a significant contribution to $D$. The fact that $D$ is much weaker when the viscous timescale for the vortices is longer indicates that there is only a weak forward cascade of energy to the dissipative scales, and that the dissipation in the fiducial simulation with standard viscosity was dominated by viscous damping of the vortices. Once again, we observe that the presence of strong vortices drastically reduces the energy injection rate into the waves, as is illustrated in the third panel of Fig.~\ref{tvol2}.
 
 \begin{figure}
  \begin{center}
          \subfigure{\includegraphics[trim=0cm 0cm 0cm 0cm, clip=true,width=0.48\textwidth]{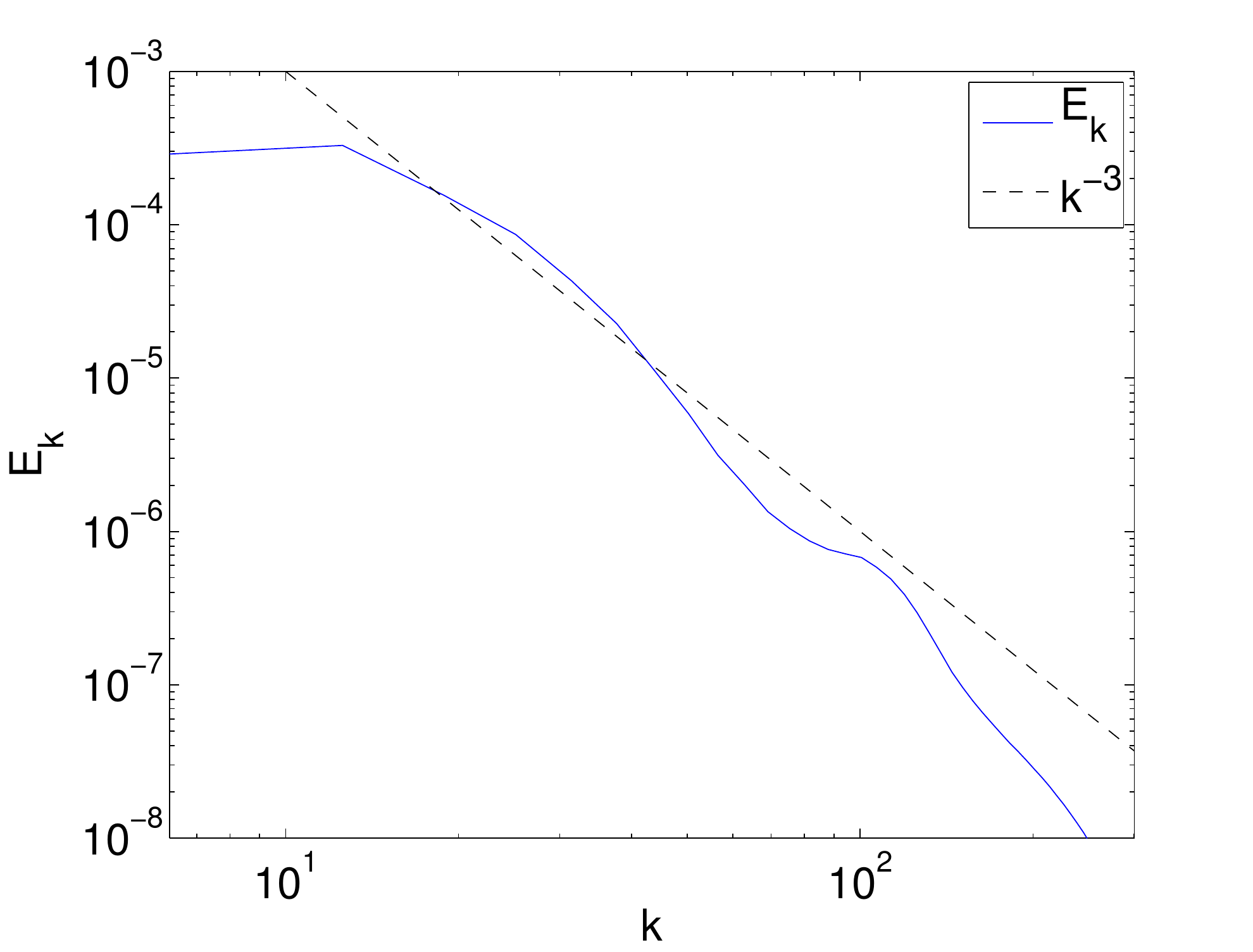}}
    \end{center}
  \caption{1D energy spectrum as a function of wavenumber for our fiducial simulation with $\alpha=4$ and $\nu_{4}=10^{-20}$. This has been computed from taking an average from $t=1000$ until the end of the simulation. The dashed line is $k^{-3}$.}
  \label{spec}
\end{figure}

We plot the 1D spherical energy spectrum, averaged over 4000 rotations, in Fig.~\ref{spec} . When the flow is dominated by vortices, the 1D spherical energy spectrum is very steep, $\sim k^{-3}$.  This is much steeper than the Kolmogorov spectrum, for example, and is consistent with a smooth velocity field (in contrast, during the transient turbulent phase at $t\sim 100$, the spectrum has a slope closer to $-5/3$).
 
This simulation illustrates that increasing the viscous timescale for the vortices strongly reduces the dissipation rate, and prevents cyclic behaviour from occurring throughout the duration of this simulation. We also confirm that whenever the flow is dominated by columnar vortices, the periodic strain is unable to efficiently excite waves. During the later stages, when a vortex grows to the size of the box, the confining effects of the horizontal boundaries will become important. However, in this simulation, the behaviour described above is seen well before the vortices grow to a sizeable fraction of the domain.

\subsection{Varying $\epsilon$ and $\Omega$: two different regimes}
\label{rotatingstrain}

The simulations described previously both had $\epsilon=0.1$, whereas we expect $\epsilon \lesssim 10^{-2}$ astrophysically. In addition, we assumed that the deformation was stationary in the inertial frame, i.e. that the secondary body was stationary, with $n=0$. Clearly this is unphysical. In this section we study how the evolution depends on $\epsilon$, as well as on the rotation rate of the deformation. Since we define our unit of time by assuming $\gamma=1$, the relevant parameter determining the importance of rotation on the flow, which is the ratio of the linear inertial wave timescale to the nonlinear advection timescale, is the Rossby number Ro $=\epsilon/\Omega$. The parameter Ro determines the relative importance of strain to rotation in the background flow. In the previous section we had Ro $=0.1$. We now look at the effects of varying this parameter through a combination of varying $\Omega$ and $\epsilon$. The parameter study described in this section used simulations with a resolution of $128^{3}$ and $\nu_{4}=10^{-18}$, in order to allow a more complete exploration of parameter space. 

The decay of the vortices seen in Fig.~\ref{tvol} is caused by viscous decay. In real stars, where viscosity is much smaller than we can simulate, the viscous decay rate should be much smaller. In order to model such a regime, we adopt hyperviscosity, since for a given resolution, hyperviscosity allows much less dissipation on large scales. Although hyperviscosity is artificial, using ordinary viscosity is also unrealistic because it requires
a viscous dissipation rate on large scales that is far too large. In addition, hyperdiffusion increases the range in $\epsilon$ over which we can study the instability (we were unable to simulate cases with the smallest $\epsilon$ considered in this section using ordinary viscosity -- the required value of $\nu$ to allow the elliptical instability to grow was too small to be simulated).

\begin{figure}
  \begin{center}
     \subfigure{\includegraphics[trim=0cm 0cm 0cm 0cm, clip=true,width=0.48\textwidth]{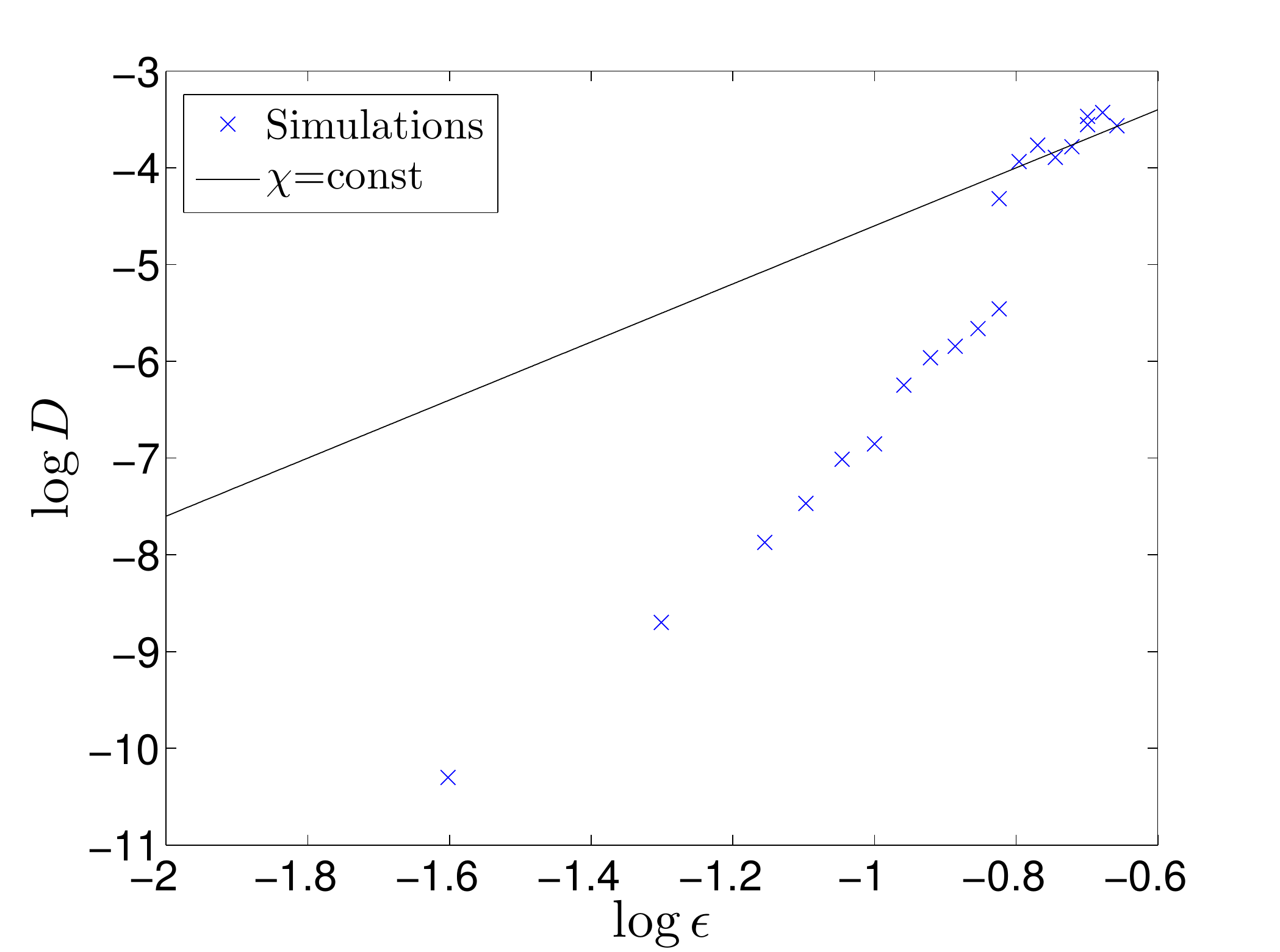}}
      \subfigure{\includegraphics[trim=0cm 0cm 0cm 0cm, clip=true,width=0.48\textwidth]{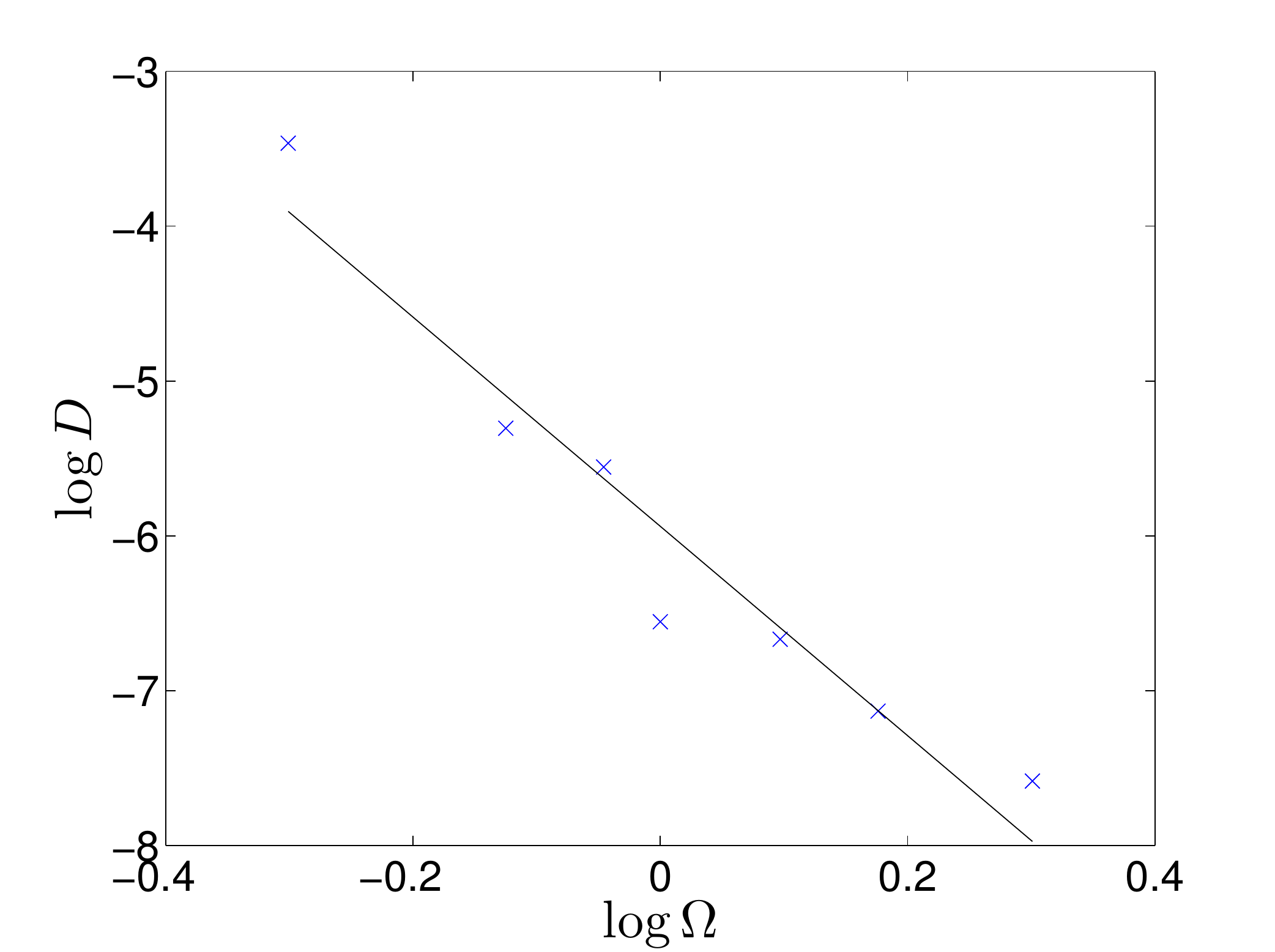}}
    \end{center}
  \caption{Top: Mean dissipation rate as a function of $\epsilon$ (when $\Omega=1,n=0$).
  A qualitative difference in the evolution is observed around $\epsilon \sim 0.15$, below which the flow produces columnar vortices, which significantly weaken the dissipation. The black solid line shows a slope of $3$ i.e., $\chi=$ const. Bottom: Mean dissipation rate as a function of $\Omega$ (when $\epsilon=0.1$).}
  \label{DissCompRo}
\end{figure}

We define the mean dissipation rate to be $(1/t)\int D(t) \mathrm{d} t$, and this is plotted in the top panel of Fig.~\ref{DissCompRo} as a function of $\epsilon$. This is the relevant quantity for astrophysical applications. If we integrate over a sufficiently long duration, this will eliminate the fluctuations in the dissipation rate, and enable the computation of a meaningful average. This is particularly important when coherent vortices are quasi-periodically formed and destroyed, since $D$ then varies by several orders of magnitude in a single simulation. Here we ensure the integrations are performed over a time interval of length $5000$, which appears sufficient to reduce the effects of the fluctuations on the average.

\begin{figure}
  \begin{center}
     \subfigure{\includegraphics[trim=1cm 14cm 1cm 1cm, clip=true,width=0.48\textwidth]{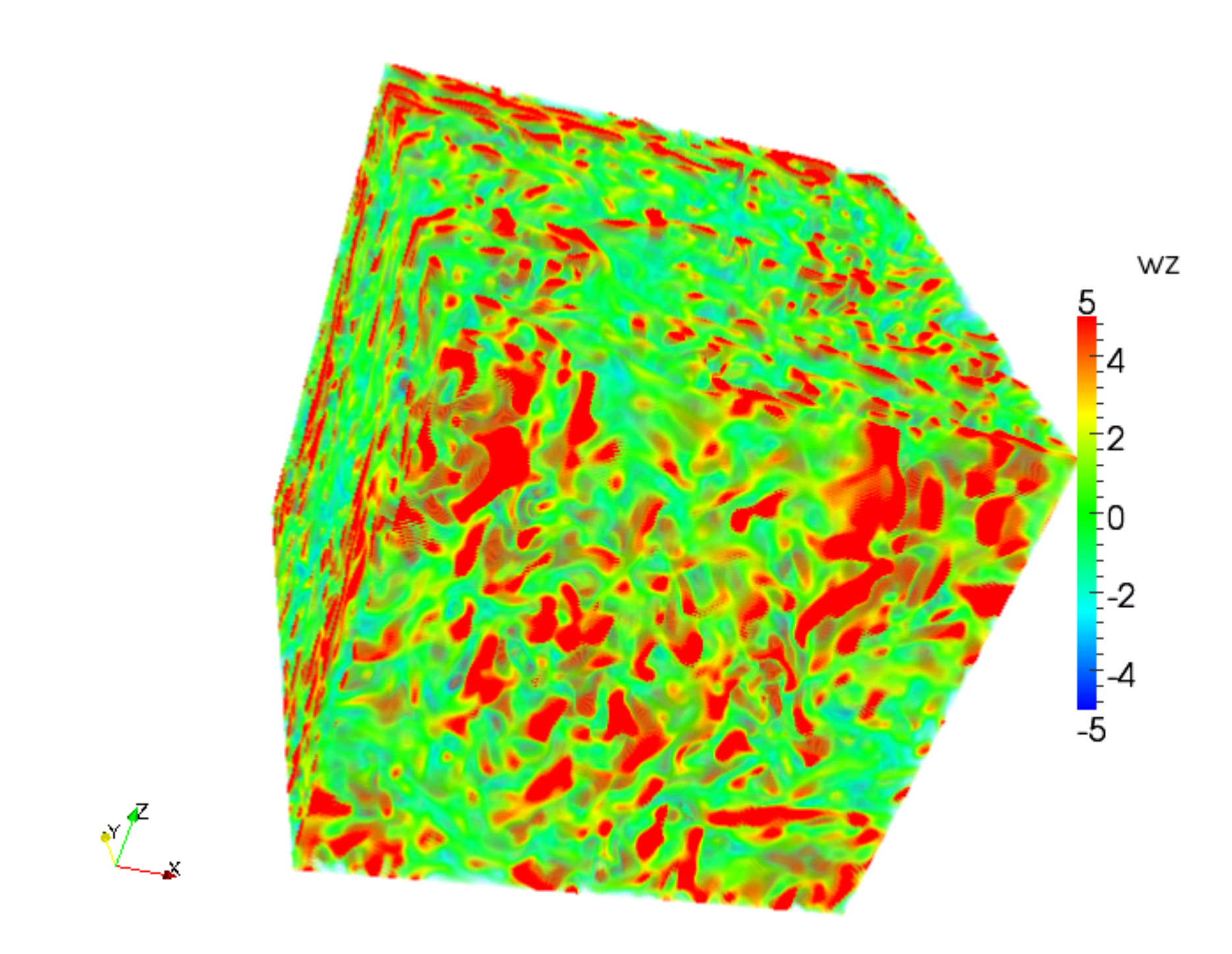}}
    \end{center}
  \caption{Volume rendering of the vertical vorticity of the flow at $t=2200$ in a simulation with $\epsilon=0.2$, with $\alpha=4$ and $\nu_{4}=10^{-18}$. This illustrates the absence of coherent vortices in the flow, and shows that the flow is more turbulent than the simulation presented in Fig.~\ref{vol}.}
  \label{Volumee0p2}
\end{figure}

The most important result obtained from varying $\epsilon$, is that when $\epsilon \gtrsim 0.15$, the evolution is qualitatively different to that observed when $\epsilon \lesssim 0.15$.  The mean dissipation rate is illustrated for several simulations with different $\epsilon$ (all with $n=0$) in the top panel of
Fig.~\ref{DissCompRo}. It is observed that there is a significant difference between simulations in which the rotation is larger but not much larger than the strain, to cases in which $\epsilon \ll \Omega$. The crossover occurs when $\epsilon \sim 0.15$, below which the flow is dominated by strong columnar vortices, which are absent when $\epsilon \gtrsim 0.15$. An illustration of the turbulence when $\epsilon=0.2$ is given in Fig.~\ref{Volumee0p2}, which shows that the flow is more turbulent than in Fig.~\ref{vol}, and does not contain strong columnar vortices. Since these are absent when $\epsilon \gtrsim 0.15$, the energy injection into the waves is sustained, and the turbulence that results is statistically steady. In this regime, the parametric driving of the waves is efficient, and the fluid responds by cascading the energy to small scales, where it can be dissipated. This results in a more efficient dissipation than in cases where the wave driving is perturbed by coherent vortices. Nevertheless, this behaviour is not relevant for the astrophysical regime, in which $\epsilon \ll 1$, where the flow develops strong columnar vortices, which inhibit the driving of the waves, and result in much weaker turbulence. 

Two different power laws are evident in Fig.~\ref{DissCompRo}, corresponding to these two regimes. For $\epsilon \lesssim 0.15$, the slope is steep, with $D\sim \epsilon^{6.5}$, whereas for $\epsilon \gtrsim 0.15$ we find $D\sim \epsilon^{3.5}$. In both of these regimes, this implies that $\chi$ is an increasing function of $\epsilon$, with $\chi \sim \epsilon^{3.5}$ or $\chi \sim \epsilon^{0.5}$, for a given $\Omega$, respectively. However, over such a short range in $\epsilon$, it is not possible to distinguish the slope for $\epsilon \gtrsim 0.15$ from that which we would predict if $\chi=$ const. We will discuss the implications in \S \ref{discussion}.

In the presence of a rotating deformation, the instability evolves similarly to the cases in which the deformation is stationary, as long as we are in the regime in which the instability is able to operate (see Fig.~\ref{rotgrowth}). This is presumably because the growth rate is not changed significantly by rotation. When Ro $\ll 1$, the flow is dominated by columnar vortices, otherwise the flow does not produce coherent vortices. We plot the mean dissipation rate as a function of $\Omega$ (for $\epsilon=0.1$) in the bottom panel of Fig.~\ref{DissCompRo}. The solid line represents a slope of $-6.5$, which reasonably represents the data.

\subsection{Aspect ratio \& resolution effects}

\begin{figure}
  \begin{center}
      \subfigure{\includegraphics[trim=0cm 0cm 0cm 0cm, clip=true,width=0.4\textwidth]{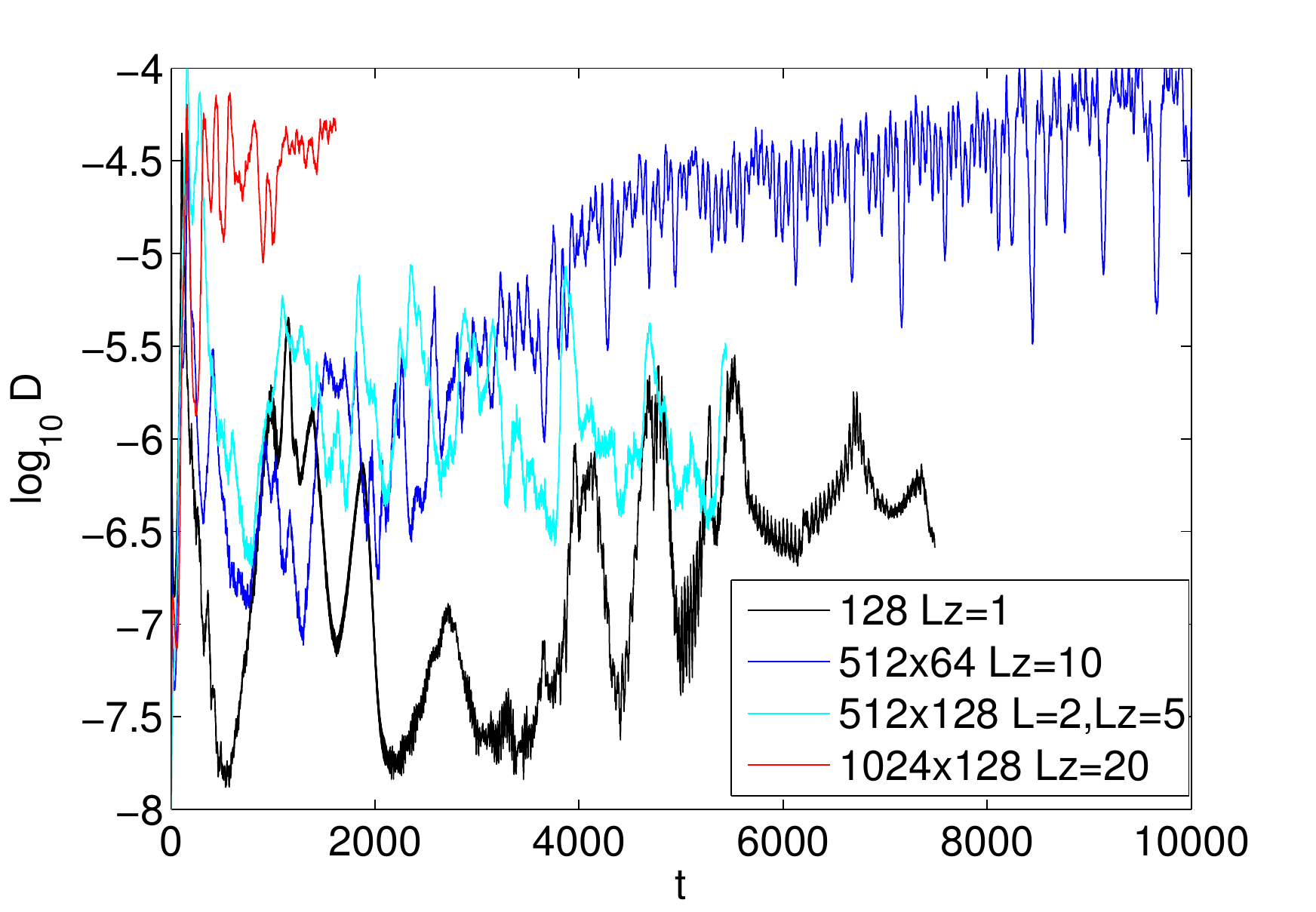}}
      \subfigure{\includegraphics[trim=0cm 0cm 0cm 0cm, clip=true,width=0.4\textwidth]{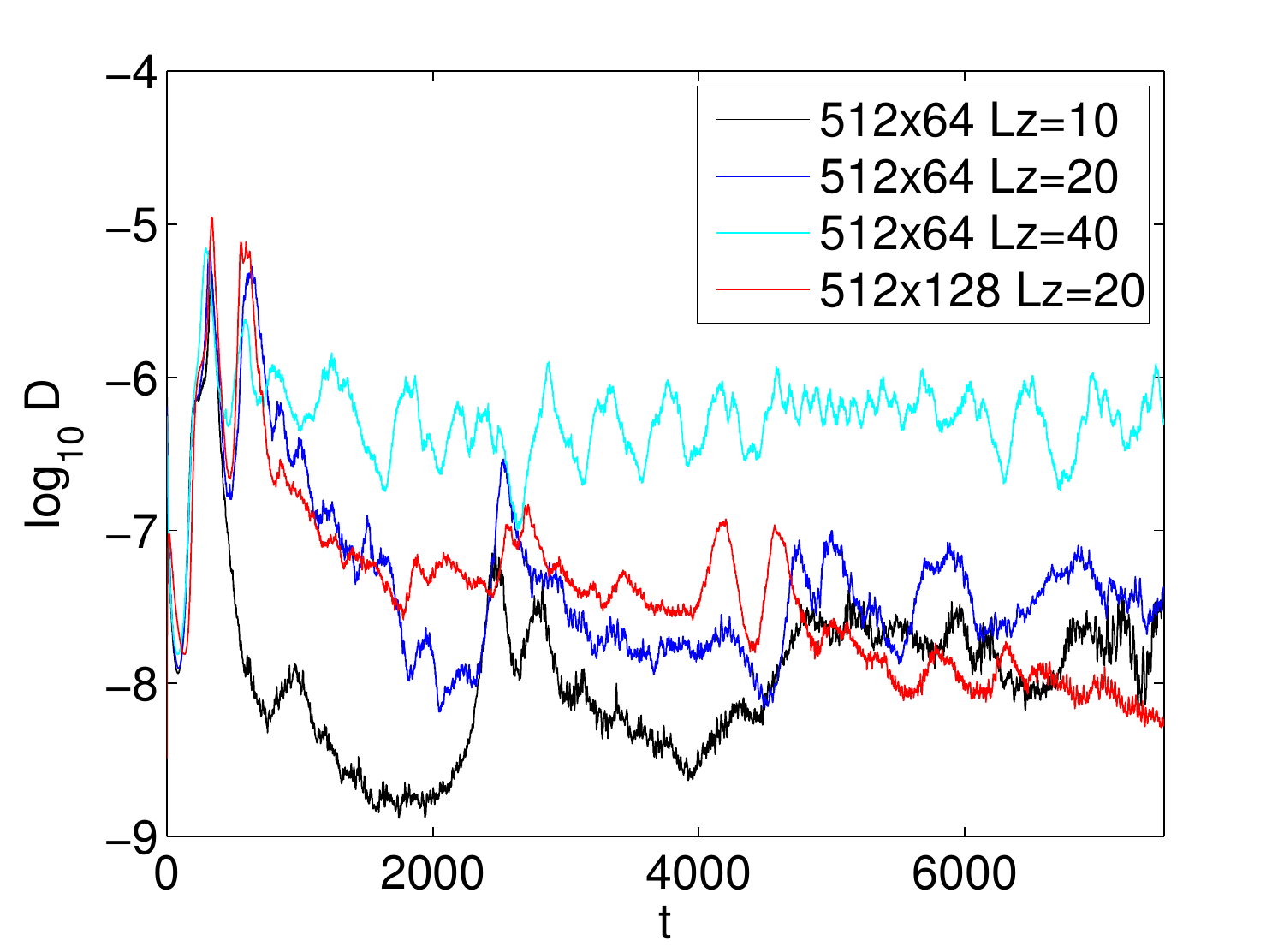}} 
    \end{center}
  \caption{Dissipation rate as a function of time for various aspect ratios for $\epsilon=0.1$ (top) and $\epsilon=0.05$ (bottom).}
  \label{Aspect}
\end{figure}

Since the flow resulting from the instability is strongly anisotropic when Ro $\ll 1$, with a preference for the formation of structures elongated along the rotation axis, it might be thought that increasing the length of the domain in the $z$-direction relative to the horizontal directions could make the flow more turbulent. This is because columnar vortices should be more unstable to long wavelength modes in the vertical direction. Increasing the aspect ratio $\beta$ increases the number of low frequency modes with small $k_{z}/k$, which might couple better to the vortices, and the absence of such modes in the unit box cases considered previously could artificially suppress the turbulence.
A similar effect is found in accretion disks, whereby a vortex becomes more unstable
when it is simulated in a taller box
\citep{Lithwick2}.

We have performed simulations with $\epsilon=0.1$ in which the aspect ratio of the box $\beta > 1$. When $\beta >1$, the amplitude of the fluctuations in the dissipation rate are somewhat reduced, and the mean is somewhat larger. However, the mean dissipation in the saturated state is found not to appreciably change until $\beta \gtrsim 10$, and the efficiency factor $\chi \ll 1$. In the top panel of Fig.~\ref{Aspect}, we have plotted the mean dissipation rate as a function of time for 4 simulations. The black curve represents a simulation with hyperviscosity, described in \S \ref{hypervis}. The others have $\beta = 10$ (blue) and $\beta=5$ with $L_{z}=5$ and $L=2$ (light blue), and finally $\beta = 20$ (red). The resolution of these simulations is $512\times 64^2$ and $512\times 128^2$ (both with $\alpha=4$ and $\nu_{4} =5 \times 10^{-15}$) and $1024\times 128^2$ (with $\nu_{4}=5\times 10^{-18}$), respectively. It would probably have sufficed to choose lower vertical resolutions than the rather high values chosen here, since the length-scales in the vertical direction tend to far exceed those in the horizontal directions. However, we wanted to ensure that the initial turbulent stages were properly resolved, and the turbulence in the initial stages is more isotropic than in the later stages. The simulation with $\beta = 10$ clearly has significantly higher dissipation than the unit box case with $\beta = 1$, differing by a factor of $10^3$, while the $\beta=20$ case is similar to the final value attained with $\beta = 10$. Visualisations of the flow indicate that the box in these large $\beta$ simulations contain a single tall cyclonic vortex, and this vortex is unstable to long wavelength modes in the vertical that could not fit into a box with $\beta = 1$. This vortex is unstable, but it is then reformed.

We can confirm that this enhanced dissipation is not a resolution issue
by comparing the blue and light blue curves in the top panel. These have the same $L_{z}$ and the same number of Fourier modes per unit length (and value of $\nu_{4}$), and only differ in the choice of  the horizontal dimension. This indicates that the reason for the much larger dissipation in the large $\beta$ cases is due to the fact that the vortices produced in these simulations are constrained by the horizontal boundaries to be tall and thin, and these vortices are more unstable. The vortices in the $\beta=10$ simulation are tall and thin, and they are constrained to be so by the horizontal boundaries, whereas the vortices produced when $\beta=5$ are not as strongly constrained in the horizontal. The picture presented here is that tall, thin vortices in the large aspect ratio simulations are much more unstable than the shorter, fatter vortices that are produced in the unit box, but this is only important once their aspect ratios exceed a critical threshold, which does not appear to be related to resolution. A possible explanation for this threshold aspect ratio above which the vortices become more unstable is the following (see also \citealt{Lithwick2}): if the mean magnitude of vorticity of these vortices is $O(\epsilon)$, these are unstable to waves with frequencies comparable with this value, corresponding with resonances with large vertical wavelength modes. From the dispersion relation for inertial waves, this requires $\beta \gtrsim O(\epsilon^{-1})$, which approximately corresponds with the threshold observed in our simulations, as presented in Fig.~\ref{Aspect}. When $\beta \gtrsim O(\epsilon^{-1})$, though at any one time the flow typically contains columnar vortices, these are strongly unstable, and an efficient and statistically steady dissipation rate is obtained. This is much larger than the values observed when $\beta \lesssim O(\epsilon^{-1})$, in which the vortices that dominate the flow are less violently unstable to large vertical wavelength modes.

Similar results are observed in the bottom panel of Fig.~\ref{Aspect}, in which the mean dissipation rate in several simulations with $\epsilon=0.05$ are presented. No appreciable difference in the dissipation rate is observed between the $\beta=10$ and $\beta=20$ cases, whereas the simulation with $\beta = 40$ has significantly higher dissipation by more than an order of magnitude. Resolution in the horizontal direction does not seem to increase the dissipation, as can be seen from the red curve, which is the same as the blue curve except we have doubled the number of grid points and reduced the hyper-viscosity to $\nu_{4}=5\times 10^{-18}$. This indicates that the dissipation rate does not appear to significantly vary with resolution. In addition, since the dominant vortices grow to the horizontal size of the box, the turbulence is much more significant for vortices that are constrained to be thin by the horizontal boundaries.

We conclude that the dissipation resulting from the elliptical instability, in the regime in which columnar vortices are formed ($\epsilon \ll 1$), is appreciable only when the aspect ratio of the domain exceeds a critical value, and when the vortices are strongly and artificially constrained by the horizontal boundaries to be tall and thin. The dissipation rate does not appear to be dependent on resolution, as far as these simulations allow us to determine. In general, the turbulence is inefficient at dissipating energy and $\chi \ll 1$, even for large aspect ratio domains.

\section{Discussion}
\subsection{Astrophysical importance of the elliptical instability}
\label{discussion}
We have performed numerical simulations of the nonlinear evolution of the elliptical instability, in order to determine its relevance as a mechanism for tidal dissipation. We have determined the efficiency factor $\chi$, and its dependence on the parameters of the problem, and we will now use this to extrapolate to the astrophysical regime.

Two different scalings for the efficiency of the dissipation are evident in Fig.~\ref{DissCompRo}. One is consistent with $D\sim 10^{-2} \epsilon^{3}$ (i.e.~$\chi \sim 10^{-2}$), when $\epsilon \gtrsim 0.15$ (the best fit slope is $D\approx0.08\epsilon^{3.5}$). The other, for $\epsilon \lesssim 0.15$, has a much steeper drop off as $\epsilon$ is decreased, with $D\approx 10^{-6}\epsilon^{6.5}$.
Note that there is significant uncertainty in these scalings, since we are determining them from a limited range of $\epsilon$.
If we believe our simulations, the latter appears to be relevant for the astrophysical regime, in which $\epsilon \ll 1$.

The application of our results to astrophysics is unclear. In our simulations, we typically observed vortices to grow until they became comparable in size to the box. A naive extrapolation of this result suggests that these vortices would occupy the entire convective region of a planet or star, and could eliminate the elliptical instability throughout. However, it remains to be seen whether these vortices would persist as the resolution is increased, far beyond what our current computational resources allow. It is possible that these vortices would become unstable to their own instabilities, thereby allowing sustained turbulence to result. Secondly, it is unclear what is the influence of periodic boundary conditions in all three directions on the evolution of the flow, when it is composed of coherent vortices. Thirdly, in convective regions of rotating fluid spheres or spherical shells, the inertial waves that may be excited by the elliptical instability are generally found to have very complicated spatial structure \citep{Gio2004,Wu2005b,Gio2007,IvanovPap2007}. It is possible that these geometrical effects would modify the dissipation resulting from the elliptical instability. These issues will be examined in global simulations of the elliptical instability, which are currently in progress.

To illustrate the relevance of the most optimistic dissipation obtained in our simulations, we can use Eqs.~\ref{tecc} and \ref{tsync} to estimate the resulting circularisation and synchronisation timescales for a typical hot Jupiter, taking $\chi=10^{-2}$. These estimates suggest that the elliptical instability could, at best, be responsible for circularising the orbits of initially eccentric hot Jupiters with $P\lesssim 2.1 $ d, and synchronising their spins if $P\lesssim 5.5$ d, within 1 Gyr. However, $\chi$ does appear to depend on $\epsilon$, which we have not included in these estimates due to our uncertainties in the scalings. If the stronger scaling found for $\epsilon \lesssim 0.15$ is appropriate, the corresponding circularisation and synchronisation periods are much shorter. This mechanism would then only result in sufficiently short evolutionary timescales to be astrophysically relevant if $P< 1 $ d.

If our local model has correctly captured the dissipative properties of the turbulence, then $\chi \ll 1$ in the astrophysical regime. This suggests that the instability could play at best a modest role in the circularisation and synchronisation of hot Jupiters. In a companion paper, we provide evidence that the addition of a weak magnetic field is able to qualitatively change the evolution when $\epsilon\ll1$, and result in a dissipation scaling as $\epsilon^{3}$, so that the optimistic estimates above may be appropriate.

\subsection{Comparison with previous work}

Experiments of the elliptical instability have been performed within a deformed elastic boundary filled with fluid (e.g.~\citealt{Lacaze2004,LeBars2007,LeBars2010}), and this has also been studied numerically by \cite{Cebron2010}. The initial instability involving the exponential growth of inertial waves is observed, which then become unstable, and the flow collapses to small-scale disorder. Once sufficient energy in the modes has been dissipated, inertial waves are again exponentially amplified, leading to cyclic behaviour. When the bulge is stationary, the ``spin-over" mode tends to be excited, which corresponds with a rigid tilt mode of the fluid (modified by boundary layers). Our local model with periodic boundary conditions is unable to capture the excitation of this mode. However, it is unclear what is the relevance of this component for tidal dissipation in fluid bodies, since it is not excited unless the bulge is stationary in the inertial frame (or alternatively, if we have a rigid boundary, and the polar axis is the middle axis). Since the experiments use an elastic boundary to contain the fluid, this will effectively impose a no-slip condition, which allows a torque to be imposed at the boundary, unlike the free boundary expected at the surface of a giant planet or star. The boundary is also ``rigid" and not a free boundary. This might overemphasise the importance of large-scale ``spin-over" modes, in which the fluid is made to rotate about a different axis than the initial rotation axis.

In this paper we avoided imposing an outer rigid boundary by using periodic boundary conditions in a small patch of the flow, and we have begun to study the asymptotic regime in which the deformation is weak, so that the growth time of the instability is much longer than the dynamical timescale. In this regime, coherent vortices are produced, which perturb the phase of the waves, preventing their coherent amplification by the strain. This behaviour may not be captured in the experiments, presumably both because of the boundary conditions, and because they do not primarily work in the $\epsilon \ll 1$ regime studied in this paper.

The elliptical instability has previously been considered in the astrophysical literature by \cite{Goodman1993}, for the problem of accretion discs in binary star systems. He found that tidally perturbed hydrodynamic accretion discs are unstable to a local linear parametric instability that excites inertial waves, which is analogous to the elliptical instability considered here. In a subsequent paper, \cite{RyuGoodman1994} studied the nonlinear evolution using 2D numerical simulations in the meridional plane, which used a shearing sheet approximation, modified to take into account the large-scale deformation of the disc. This approach is somewhat similar to ours, in principle, though the details differ. They found that the linear instability resulted in turbulence and continual energy dissipation, implying a secular tidal torque. Their method would not be able to capture columnar vortices aligned with the rotation axis because they neglected variations of the fluid variables in the azimuthal direction. However, we might not expect such vortices to be produced in their case because the elliptical deformation is not much weaker than the rotation, and their problem contains the additional effect of Keplerian shear. Our simulations are fully 3D, and we study a wide range of ratios of strain to rotation, finding sustained turbulence whenever this ratio is larger than $\sim 0.15$, but not when it is much smaller. It would be of interest to study their problem again in 3D, perhaps using a similar method to ours, using an appropriate background flow that is linear in the coordinates.

\section{Conclusion}

In this paper we have presented an initial study into the nonlinear evolution of the elliptical instability in a fluid tidally deformed planet or star, with the aim to determine its relevance for tidal dissipation. We adopted an idealised local model, whose simplicity permits high-resolution numerical simulations using a pseudo-spectral method. This allowed us to study in detail the effects of a time-dependent elliptical deformation on the evolution of perturbations to a uniformly rotating unstratified incompressible fluid.

We found that, in the astrophysical regime, in which the elliptical strain is much weaker than the rotation of the fluid, the flow organises itself into strong columnar vortices aligned with the axis of rotation (``Proudman-Taylor columns"), in the presence of which, subsequent excitation of waves is predominantly suppressed. Therefore, when the ellipticity is small, the instability saturates by suppressing the driving mechanism of the instability. This is presumably because the phases of any potentially unstable modes are detuned, so that they cannot be coherently driven, resulting in much less dissipation than we might expect if the wave driving is sustained. On the other hand, if the strain is not much weaker than the rotation of the fluid, sustained turbulence is possible, and the dissipation rate is much higher -- however, this is not observed in our simulations in the astrophysically relevant regime of small ellipticity.

Our main result is that it appears, from our local hydrodynamical model, that the turbulence generated through the nonlinear evolution of the elliptical instability is unlikely to provide an explanation for the required level of tidal dissipation in gaseous planets and close-binary stars (the latter was first proposed as a possibility by \citealt{Rieutord2004}). However, our model neglects a number of effects, including magnetic fields, realistic geometry, and the additional presence of turbulent convection. Whether these can revive the elliptical instability as a plausible mechanism for tidal dissipation remains to be seen.

\section*{Acknowledgments}
We would like to thank the referee for a prompt and careful reading of the manuscript, whose report helped to clarify several points in the paper. YL acknowledges the support of NSF grant AST-1109776.

\setlength{\bibsep}{0pt}
\bibliography{tidbib}
\bibliographystyle{mn2e}
\end{document}